\documentclass[fleqn,usenatbib]{mnras}

\usepackage{newtxtext,newtxmath}

\usepackage[T1]{fontenc}

\DeclareRobustCommand{\VAN}[3]{#2}
\let\VANthebibliography\thebibliography
\def\thebibliography{\DeclareRobustCommand{\VAN}[3]{##3}\VANthebibliography}

\usepackage[utf8]{inputenc} 
\usepackage[T1]{fontenc}    
\usepackage{hyperref}       
\usepackage{url}            
\usepackage{nicefrac}       
\usepackage{microtype}      
\usepackage{xcolor}         

\usepackage{amsmath,amsfonts,bm}

\def\eqref#1{equation~\ref{#1}}

\def\1{\bm{1}}

\def\vmu{{\bm{\mu}}}

\def\vx{{\bm{x}}}
\def\vy{{\bm{y}}}

\def\mC{{\bm{C}}}

\def\mF{{\bm{F}}}

\def\mX{{\bm{X}}}

\DeclareMathAlphabet{\mathsfit}{\encodingdefault}{\sfdefault}{m}{sl}
\SetMathAlphabet{\mathsfit}{bold}{\encodingdefault}{\sfdefault}{bx}{n}

\DeclareMathOperator{\Tr}{Tr}

\usepackage{hyperref}
\usepackage{url}

\usepackage{subcaption}
\usepackage[]{caption}
\usepackage{cleveref}

\usepackage{lastpage}

\usepackage{graphicx}
\graphicspath{ {./images/} } 

\usepackage{amsfonts}
\usepackage{amsmath}

\usepackage{bm}  
\usepackage{mathtools}
\usepackage{dsfont}

\usepackage{courier}

\usepackage{multicol}

\usepackage{booktabs}
\usepackage{multirow}

\usepackage{footmisc}

\title[RGZ: SSL for radio galaxy classification]{Radio Galaxy Zoo: Using semi-supervised learning to leverage large unlabelled data-sets for radio galaxy classification under data-set shift}

\author[Slijepcevic et~al.]{
Inigo V.~Slijepcevic$^{1}$\thanks{E-mail: inigo.slijepcevic@postgrad.manchester.ac.uk},
Anna M.~M.~Scaife,$^{1, 2}$
Mike Walmsley,$^{1}$
Micah Bowles,$^{1}$
O.~Ivy Wong,$^{3,4,5}$
\newauthor
Stanislav~S.~Shabala$^{6,5}$
and
Hongming Tang$^{7}$
\\
$^{1}$ Department of Physics and Astronomy, University of Manchester, Manchester, UK\\
$^{2}$ The Alan Turing Institute, Euston Road, London, NW1 2DB, UK\\
$^{3}$ CSIRO Space \& Astronomy, PO Box 1130, Bentley, WA 6102, Australia \\
$^{4}$ ICRAR-M468, University of Western Australia, Crawley, WA 6009, Australia \\
$^{5}$ ARC Centre of Excellence for All Sky Astrophysics in 3 Dimensions (ASTRO 3D), Australia \\
$^{6}$ School of Natural Sciences, Private Bag 37, University of Tasmania, Hobart, TAS 7001, Australia \\
$^{7}$Department of Astronomy, Tsinghua University, Beijing 100084, China\\
}

\date{Accepted XXX. Received YYY; in original form ZZZ}

\pubyear{2021}

\begin{document}
\label{firstpage}
\pagerange{\pageref{firstpage}--\pageref{lastpage}}
\maketitle

\begin{abstract}
In this work we examine the classification accuracy and robustness of a state-of-the-art semi-supervised learning (SSL) algorithm applied to the morphological classification of radio galaxies. We test if SSL with fewer labels can achieve test accuracies comparable to the supervised state-of-the-art and whether this holds when incorporating previously unseen data. We find that for the radio galaxy classification problem considered, SSL provides additional regularisation and outperforms the baseline test accuracy. However, in contrast to model performance metrics reported on computer science benchmarking data-sets, we find that improvement is limited to a narrow range of label volumes, with performance falling off rapidly at low label volumes. Additionally, we show that SSL does not improve model calibration, regardless of whether classification is improved. Moreover, we find that when different underlying catalogues drawn from the same radio survey are used to provide the labelled and unlabelled data-sets required for SSL, a significant drop in classification performance is observered, highlighting the difficulty of applying SSL techniques under dataset shift. We show that a class-imbalanced unlabelled data pool negatively affects performance through prior probability shift, which we suggest may explain this performance drop, and that using the Frechet Distance between labelled and unlabelled data-sets as a  measure of data-set shift can provide a prediction of model performance, but that for typical radio galaxy data-sets with labelled sample volumes of $\mathcal{O}(10^3)$, the sample variance associated with this technique is high and the technique is in general not sufficiently robust to replace a train-test cycle. 
\end{abstract}



\section{Introduction}
\label{subsec:sciencegoal}


Radio galaxies are a subset of active galactic nuclei (AGN) that typically exhibit a pair of roughly symmetric jets that are usually pointed in opposite directions, although high ram pressure can cause some examples to exhibit a "bent-tailed" structure \citep{Mguda2014}. The jets are powerful emitters across the electromagnetic spectrum and their radio emission is dominated by synchrotron emission from ultra-relativistic electrons \citep{hardcastle2020}. Historically, radio galaxy morphologies have been assigned to various categories. The most persistent amongst these is the Fanaroff Riley classification \citep{Fanaroff1974}, with radio galaxies split into the Fanaroff Riley type I (FRI) and Fanaroff Riley type II (FRII) categories based on the distance between the brightest point on each lobe as a proportion of the total length of the source \citep{Fanaroff1974}. FRI (FRII) sources have lower (higher) brightness lobes, whose brightness typically decrease (increase) as distance to the center increases. 



 
Although progress has been made in relating the two FR classes to the dynamics and energetics of the sources \citep[e.g.][]{Ineson2017, Saripalli2012, Turner2015ENERGETICSNUCLEI, Hardcastle2018AGalaxies}, our understanding of the causal relationship between a source's FR classification and physical environment/properities is incomplete with a number of outstanding questions that still remain. In particular many existing studies are tied to strongly flux-density-limited samples of radio galaxies with differing redshift distributions, which impose difficult selection biases for conclusive population studies \citep{hardcastle2020}. More sensitive surveys across a wider range of wavelengths have revealed a more morphologically diverse population of observed examples. These populations challenge existing paradigms, in particular those tying morphological FR classification to radio luminosity \citep[e.g.][]{Mingo2019}. Furthermore, increased sensitivity in the radio region allows us to observe more galaxies, with some examples such as remnant galaxies \citep{Brienza2016, Murgia2011, Brienza2016a}, which are though to be FRI/FRII descendants, fitting into the existing paradigm. However, there are also examples that do not fit well into the FRI/II schema, such as hybrid radio galaxies \citep{Gopal-Krishna2000}. There is further ambiguity with a morphological split between compact and extended sources \citep{Miraghaei2017}; whether compact sources are a separate class of radio galaxy or an "FR0" precursor to extended FRI/II sources is still unknown \citep{Baldi2015}. This raises the (open) question of whether FRI/FRII classification is an optimal classification scheme, particularly if we are using it as ground truth to train a supervised model.


When building samples of radio galaxies with which to test models linking morphology to physical characteristics, an additional complication is the volume of data produced by modern radio telescopes. The Rapid ASKAP continuum survey \citep{racs2020} has already detected three million extended sources and the upcoming Evolutionary Map of the Universe (EMU) survey is expected to detect 70 million radio sources \citep{norris2006, Norris2011EMU:Universe, Norris2021TheSurvey}. Such large data volumes are not compatible with the \emph{by eye} approaches to classification that have been widely used historically. This has resulted in an increased development and utilisation of automated detection and classification. Consequently, machine learning methodologies have continued to gain traction as automated astronomical image classification tools. In particular, for future surveys with the Square Kilometre Array (SKA), which is expected to produce an unprecedented volume of data: $\sim$ 1 petabyte of image data per day \citep{Hollitt2016}, detecting up to 500 million new radio sources \citep{Prandoni}, machine learning approaches that can effectively process huge data volumes will become essential.

In particular, Convolutional Neural Networks (CNNs) have been successfully applied to image-based classification of radio galaxies. The original work in this field is \cite{Aniyan2017} who classified radio galaxies into FRI and FRII type objects; other works have also incorporated object detection as well as classification  \citep[e.g.][]{Wu2019,Wang2021ResearchYolov5}, and attempts to use more novel techniques such as capsule networks \citep{Lukic2018}, attention gating \citep{Bowles2021,Wang2021ResearchYolov5} and group-equivariant networks \citep{Scaife2021} to help improve performance and interpretability have also been implemented. \cite{Mohan2022QuantifyingClassification} demonstrated that a variational inference based Bayesian deep-learning approach could be used to provide calibrated posterior uncertainties on individual radio galaxy classifications. Alternatives to CNN-based approaches for classification of radio galaxies include \cite{ntwaetsile2021} who extracted Haralick features from images as a rotationally-invariant descriptor of radio galaxy morphology for input to a clustering analysis, and \cite{Sadeghi2021Morphological-basedMoments} who calculated image moments to implement a support vector machine approach to radio galaxy classification.
\citet{Becker2020} provides a comprehensive survey of current supervised machine learning techniques.

In the context of upcoming radio surveys, it is important to know how much labelling is required to achieve good classification results. Due to the high cost of labelling, reducing the size of the required labelled data-set is beneficial and will allow us to deploy working models earlier. We expect comparative model performance to be (mostly) constant across archival and new data-sets: we can therefore test different algorithms and models on the MiraBest data-set, and expect the results to hold for new surveys.

Currently, the archival data-sets available for training radio galaxy classifiers are of comparable size to many of those used in computer vision \citep[e.g. CIFAR;][]{Krizhevsky2009}, with around $10^5$ samples available). However, a fundamental difference is domain knowledge needed for creating labelled data-sets, which has a much higher cost for radio astronomy data. As a result of this, \emph{labels} are sparse in radio galaxy data-sets, with labels only included for a small fraction of data. Labelled catalogues of radio galaxies contain of order $10^3$ objects and the largest publicly available FR-labelled machine learning dataset of radio galaxies is MiraBest \citep{Miraghaei2017, Porter2020MiraBestDataset}, which has 1256 samples, orders of magnitude lower than the number of unlabelled images in its originating sky survey \citep[FIRST;][]{becker1995}. However, while labelled radio galaxy datasets are in the low data regime, we note that they are still typically significantly larger than those associated with one- or few-shot learning; although the potential of such approaches to radio galaxy classification  has been explored by \citet{Samudre2022Data-efficientGalaxies}.

Data augmentation using flips and rotations of the training data as originally proposed in \cite{Aniyan2017} for radio galaxy classification is widely used to mitigate overfitting due to this lack of (labelled) data.It has been shown that a well thought out augmentation strategy is crucial to performance in the low data regime for radio galaxy classification \citep{Maslej-Kresnakova2021MorphologicalTechniques}, and the potentially negative impact of unprincipled data augmentation, particularly in the case of Bayesian deep-learning approaches to radio galaxy classification, has been highlighted by \cite{Mohan2022QuantifyingClassification}. Therefore we note that regardless of the learning paradigm, augmentation strategy will remain an important variable in model training.

However, when adapting any computing technique for use in science, we must ask the question: is the algorithm we are using designed for our use case? How closely do our inputs match those used in the computer science literature?

It is clear that supervised learning is \textit{not} designed for data-sets limited by the number of available labels and a large proportion of (useful) unlabelled data, as these data samples are simply discarded. Therefore, in order to achieve optimal results there are two options: (i) label more data, or (ii) use an algorithm capable of leveraging the population of unlabelled samples. We focus on the second approach as option (i) will require more and more human labelling as data-set and model sizes increase. Despite some success with crowd-sourcing labels \citep{Banfield2015}, manual labelling of radio galaxies through citizen science 
is unlikely to scale to the rate of data output from telescopes such as the SKA, as non-expert labelling usually requires large group consensus to accurately label samples. It therefore seems obvious that an algorithm able to incorporate information from \textit{unlabelled} data into its predictions while still making maximal use of any available labels is desirable in the case of radio galaxy classification.

In this work we investigate whether it is possible to achieve performance comparable to the current supervised state of the art but with many fewer labels by using semi-supervised learning (SSL). In doing so, we isolate and test the effect of real world complexities in the radio astronomy data such as non-identically-distributed labelled and unlabelled datasets, unknown class imbalance in the unlabelled data, and the choice of labelled data-set. Furthermore, we present some generalised diagnostics to predict the performance of SSL algorithms to a new problem in a computationally efficient manner.

The structure of the paper is as follows: in Section~\ref{sec:ssl} we introduce the semi-supervised learning paradigm, review its previous applications in astronomy and highlight the challenges arising from data structure differences in this field; in Section~\ref{sec:data} we introduce the datasets being used in this work and their processing; in Section~\ref{sec:fixmatch} we introduce the SSL algorithm being used in this work and describe its implementation and training\footnote{Code can be found at \url{https://github.com/inigoval/fixmatch}.}; in Section~\ref{sec:results} we present the results of SSL for radio galaxy classification under various data conditions; in Section~\ref{sec:disc} we discuss these results in the context of the radio galaxy classification problem addressed in this work; and in Section~\ref{sec:conclusion} we draw our conclusions.  

\section{Semi-supervised learning}
\label{sec:ssl}

Semi-supervised learning refers to a set of solutions designed to leverage unlabelled data when a small labelled data-set is available. A full theoretical overview of semi-supervised learning is beyond the scope of this paper, but can be found in \citet{ChapelleSSL}. Unlabelled data can provide the classifier with extra information about the data manifold while also regularising the model and mitigating overfitting in the low (labelled) data regime. The model is fed a set of image-label pairs $(\vx_{\rm l}, \vy_{\rm l}) \in \mX_{\rm l}$ along with a (usually larger) set of unlabelled images $\vx_{\rm u} \in \mX_{\rm u}$. The goal is to predict labels for a set of reserved test samples, $\vx_{\rm test} \in \mX_{\rm test}$, or a new unseen dataset. 

Semi-supervised learning (SSL) has an extensive literature with many different approaches that achieve good results. It is usually desirable for models to be consistent when making predictions on perturbations/transformations of the same image, which partly motivates the widespread use of data augmentation during supervised training \citep{Chen2020}. This idea has also been adopted in semi-supervised learning, with many algorithms implementing some form of \textit{consistency regularization} which penalises the classifier purely based on the consistency of its predictions given perturbations (augmentations) of the image, rather than the correctness of the prediction (which is unknown for unlabelled data). For example, \citet{Tarvainen2017MeanResults} use an exponential moving average of previous model weights to enforce consistent predictions, while  \citet{Miyato2019VirtualLearning} take a different approach, using adversarial perturbations to force consistency in all directions around each data point, smoothing the decision manifold and pushing it away from the data. Alternatively, \textit{pseudo labelling} involves propagating "soft" labels predicted by the model onto unlabelled data-points when the model has high confidence \citep{Lee2013Pseudo, Pham2021Meta}. The FixMatch algorithm \citep{Sohn2020FixMatch:Confidence} combines consistency regularization and pseudo-labelling by propagating the pseudo-label of a weakly augmented image onto a strong augmentation of the same image. FixMatch is the focus of Section~\ref{sec:fixmatch}.

\subsection{Applications of semi-supervised and self-supervised learning in astronomy}
\label{sec:sslastro} 

SSL has been used for a broad range of applications in astronomy. Examples include \citet{Marianer2021ASources} who focus on outlier detection in gravitational wave data, \citet{Richards2012Semi-supervisedClassification} apply SSL to photometric supernova classification, and \citet{Hayat2020EstimatingLearning} who improve state of the art galactic distance estimations from images using SSL.

Perhaps the most relevant SSL applications for this work are in image classification, where the most popular approach has leveraged unsupervised models to learn a structured representation with unlabelled data first (``pretraining''), before fine-tuning (usually with labels) at the end, which is common in computer vision \citep{Chen2020}. For example, \citet{Ma2019} successfully trained an auto-encoder to create a latent representation for radio galaxies and then fine-tuned this model in the standard supervised manner with all available labels and a cross-entropy loss, to improve model accuracy.
Cutting edge contrastive learning methods have been used for the pretraining stage in galaxy morphology \citep[][]{Hayat2021Self-SupervisedImages} and for gravitational lens/non-lens \citep[][]{Stein2021} classification, with promising results indicating that this technique may work in the radio astronomy domain as well.

Self-organising maps have also been used to generate useful representations for radio data, from which classes can be inferred. This has been used successfully for identifying rare and unusual object morphologies \citep{Galvin2020, Ralph2019}. However, this technique is unable to use existing labels. Furthermore, experts are required to label a set of "prototypical" samples generated by the model, which may be large in number or have mixed features. Generative Adversarial Networks (GANs; a self-supervised generative architecture) have been used for survey-to-survey image translation with some success \citep[e.g.][]{Schawinski2017}. However, as described in \cite{Glaser2019RadioGANNetworks}, the technique is unlikely to work in the case of low to high resolution translation (so-called "super-resolution") in the case of radio astronomy, due to the nature of interferometric measurements.

Unsupervised morphological classification is attempted in \cite{Spindler2020} with a variational deep embedder. The data is clustered in the latent space of the model, which can be looked at by eye to interpret the morphological differences between clusters. While this is a worthwhile approach for separating data into meaningful groups, the cluster ``labels'' (i.e. FRI/FRII classification scheme in our case) cannot be chosen a priori, which we may require if we want to use scientifically defined categories (e.\,g. FRI/II). Furthermore, there can be some inter-cluster overlap, and we cannot control which properties the model uses most of its capacity on, resulting in scientifically unimportant parts of the image (e.\,g. secondary sources) taking up much of the model's attention.


\subsection{Data structure challenges}
\label{sec:datastructure}

\begin{figure}
\centering
\begin{subfigure}[t]{.23\textwidth}
    \centering
    \includegraphics[width=0.9\linewidth]{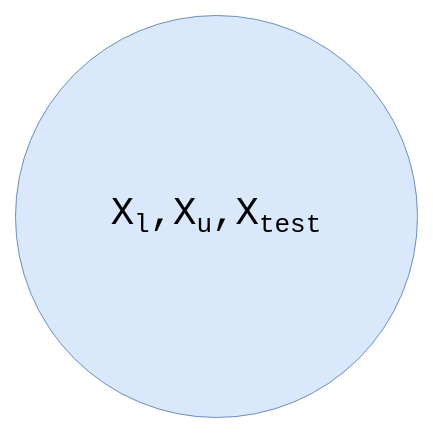}
        \caption{} 
        \label{subfig:sslset}
\end{subfigure}%
\begin{subfigure}[t]{.23\textwidth}
    \centering
    \includegraphics[width=0.9\linewidth]{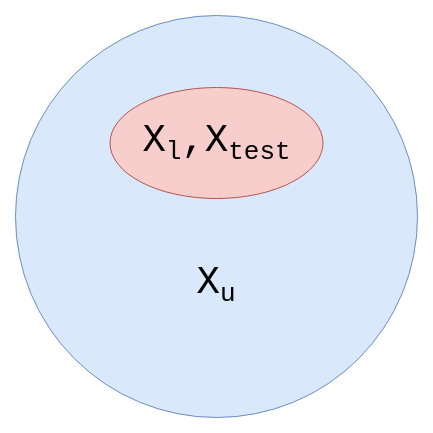}
        \caption{}
        \label{subfig:domainexpansion}
\end{subfigure}
\caption{Visual representation of data structure differences between standardised machine learning benchmark datasets and real world radio astronomy datasets. (a) Data structure in the majority of the semi-supervised learning literature, and (b) data structure in radio astronomy.}
\label{fig:datasets}
\end{figure}

State-of-the-art SSL algorithms achieve impressive accuracies on standard benchmarking data-sets with few labels: 97.13 \% on CIFAR-10 \citep{Krizhevsky2009} with 4000 labels using LaplaceNet \citep{Sellars2021LaplaceNet:Classification}, 97.64 \% on SVHN \citep{Netzer2011} with 1000 labels using Meta Pseudo-Labels \citep{Pham2021Meta} and 94.83 \% on STL-10 \citep{Coates2011AnLearning} with FixMatch. However, less work has been done in assessing the robustness of SSL to real world datasets, which may include unclean, covariate-shifted or prior-probability-shifted data, out-of-distribution unlabelled data or even simply varying proportions of labelled/unlabelled data. \cite{Oliver2018RealisticAlgorithms} give a detailed analysis of the shortcomings of the SSL literature in the context of real world applications. 

As well as the class ambiguity discussed in Section~\ref{subsec:sciencegoal}, astronomical data present different challenges to those explored in the SSL literature. In the SSL literature, it is widely assumed that $\vx_{\rm test}$, $\vx_{\rm u}$ and $\vx_{\rm l}$ are all drawn from the same distribution, as illustrated in Figure~\ref{subfig:sslset}. In the literature this holds as labels are typically discarded from a dataset to mimic an unlabelled pool, ensuring that there is no difference other than size between $\mX_{\rm u}$ and $\mX_{\rm l}$. However, in astronomy and indeed many applications in observational science, $\mX_{\rm u}$ contains previously `unseen' observations such that $\mX_{\rm u}$ and $\mX_{\rm l}$ are in general \textit{not} identically distributed. This causes \textit{covariate shift} between the unlabelled and labelled data. The distribution of labels $p_u(y)$ and $p_l(y)$ are also not identical in general, causing prior probability shift \citep{datasetshiftbook}. Depending on the specific differences in a given application, this can cause problems for SSL algorithms not designed to deal with this kind of dataset shift.


Of particular interest in astronomy is the problem of biased sampling of our training data. As a result of observational, instrumental and intrinsic effects which skew towards selecting specific data such as particular ranges of flux density (brightness) and redshift (distance), the samples chosen for labelling are chosen in a biased way. While this is unavoidable when observing phenomena driven by complex natural processes, we must be wary of how it will affect our model performance. Observational data can also become corrupted due to image processing errors, for example during (de)compression \citep[][]{Ciprijanovic2021RobustnessStudies}, which might lead to polluted unlabelled data. This makes it hard to how well our labelled data catalogues represent all observed data, with the consequent effects on model performance not always clear a priori. Specific examples of how this is being addressed for machine learning applications in astronomy include the use of Gaussian process modelling to improve data augmentation in photometric classification, making the training data more representative of test data \citep{Boone2019Avocado:Augmentation} and in galaxy merger classification where domain adaptation techniques have also been explored \citep{Ciprijanovic2020DomainMergers}. In both of these works, the authors are tackling dataset shift between the labelled and test data.

\citet{Cai2021} use pretraining with the SwAV algorithm \citep[a type of contrastive learning;][]{Caron2020UnsupervisedAssignments} and fine-tuning with FixMatch (see Section~\ref{sec:fixmatch} for details) to improve generalisation on distribution-shifted unlabelled data. Their results imply that techniques using consistency regularisation are more robust to dataset shift than domain-adaptation-based approaches, and this motivates our choice of algorithm in this work. 

\section{Data}
\label{sec:data}

We recreate the ideal SSL case shown in Figure~\ref{subfig:sslset} to test whether SSL methods can help radio galaxy classification when little data is available and the data distribution is consistent between $\mX_{\rm l}$ and $\mX_{\rm u}$. To do this, we use a stratified subset of the MiraBest data-set \citep{Miraghaei2017, Porter2020MiraBestDataset}, as our possibly-labelled dataset $\mX_l$, and discard labels from a proportion of the dataset to create our unlabelled dataset $\mX_{\rm u}$. For the more realistic case shown in Figure~\ref{subfig:domainexpansion}, where our unlabelled data is drawn from a different distribution to our labelled data, we use data selected from the Radio Galaxy Zoo Data Release 1 (RGZ DR1) catalogue (Wong et al. in prep) as $\mX_{\rm u}$.

\subsection{MiraBest}
\label{sec:mirabest}

The MiraBest machine learning data set \citep{Porter2020MiraBestDataset} consists of 1256 images of radio galaxies pre-processed for deep learning tasks. The data-set was constructed using the sample selection and classification described in \cite{Miraghaei2017}, who made use of the parent galaxy sample from \cite{Best2012OnProperties}. Optical data from data release 7 of the Sloan Digital Sky Survey \citep[SDSS DR7;][]{sdssdr7} was cross-matched with NRAO VLA Sky Survey  \citep[NVSS;][]{condon1998} and Faint Images of the Radio Sky at Twenty-Centimeters  \citep[FIRST;][]{becker1995} radio surveys. Parent galaxies were selected such that their radio counterparts had an active galactic nucleus (AGN) host rather than emission dominated by star formation. To enable classification of sources based on morphology, sources with multiple components in either of the radio catalogues were considered. 

The morphological classification was done by visual inspection at three levels: (i) The sources were first classified as FRI/FRII based on the original classification scheme of \citet{Fanaroff1974}. Additionally, 35 \emph{Hybrid} sources were identified as sources having FRI-like morphology on one side and FRII-like on the other. Of the 1329 extended sources inspected, 40 were determined to be unclassifiable. (ii) Each source was then flagged as `Confident' or `Uncertain' to represent the degree of belief in the human classification and, although this qualification was not extensively explained in the original paper, \citet{Mohan2021WeightClassification} have shown that it is correlated with model posterior variance over the dataset. (iii) Some of the sources which did not fit exactly into the standard FRI/FRII dichotomy were given additional tags to identify their sub-type. These sub-types include 53 Wide Angle Tail (WAT), 9 Head Tail (HT) and 5 Double-Double (DD) sources. To represent these three levels of classification, each source was given a three digit identifier as shown in Table~\ref{tab:digits_mirabest}.

\begin{table}
[!t]
\centering
\caption[Short table caption.]{Three digit identifiers for sources in \citet{Miraghaei2017}}
	\begin{tabular}{llll}
    Digit 1 & Digit 2 & Digit 3 \\
    
    \hline
    
    0: FRI & 0: Confident  &  0: Standard \\
    1: FRII & 1: Uncertain & 1: Double Double \\
    2: Hybrid &         & 2: Wide Angle Tail \\ 
    3: Unclassifiable & & 3: Diffuse \\
    &  &                   4: Head Tail \\ 
	\end{tabular}
   \label{tab:digits_mirabest}
\end{table}

\begin{table}

\centering
\caption[MiraBest Class-wise Composition]{MiraBest Class-wise Composition}
	\begin{tabular}{cccc}
	\hline
    Class & Confidence & No. \\
    \hline
    \multirow{2}{2em}{FRI}  & Confident & 397\\
     & Uncertain & 194\\
    \hline
    \multirow{2}{2em}{FRII}  & Confident & 436\\
     & Uncertain & 195\\
    \hline
    \multirow{2}{2em}{Hybrid}  & Confident & 19\\
     & Uncertain & 15\\
    \hline
    
	\end{tabular}
   \label{tab:mb_classes}
\end{table}

To ensure the integrity of the ML data set, the following 73 objects out of the 1329 extended sources identified in the catalogue were not included: (i) 40 unclassifiable objects; (ii) 28 objects with extent greater than the chosen image size of $150\times150$ pixels; (iii) 4 objects which were found in overlapping regions of the FIRST survey; (iv) 1 object in category 103 (FRII Confident Diffuse). Since this was the only instance of this category, it would not have been possible for the test set to be representative of the training set. The composition of the final data set is shown in Table~\ref{tab:mb_classes}. We do not include the sub-types in this table as we do not consider their classification in this work.

\subsection{Radio Galaxy Zoo}
\label{sec:rgz}

The Radio Galaxy Zoo (RGZ) project is an online citizen science project that is aimed at the morphological classification of extended radio galaxies and the identification of host galaxies \citep{Banfield2015}. The online RGZ programme operated for $\sim$5.5 years between December 2013 and May 2019.  Within this operational period, over 2.2 million independent classifications were registered for over 140,000 subjects.  The RGZ subjects inspected by citizen scientists consist of images derived from the FIRST survey \citep{becker1995} and the ATLAS survey \citep{norris2006}, overlaid onto WISE $W1$ (3.5~$\mu$m) infrared images.  The RGZ dataset used in this paper is from the RGZ Data Release 1 catalogue (RGZ DR1; Wong et al., in prep).

The RGZ DR1 catalogue contains approximately 100,000 source classifications with a user-weighted consensus fraction (consensus level) that is equal to or greater than 0.65.   Within the catalogue, 99.2\% of classifications used radio data from the FIRST survey \citep{becker1995}, and the remainder used data from the ATLAS survey \citep{norris2006}.   The largest angular size (LAS) of each source in the RGZ DR1 is estimated by measuring the hypotenuse of a rectangle that encompasses the entire radio source at the lowest radio contour \citep{Banfield2015}. This method is generally reliable if the radio components of a source are correctly-identified and the source components are distributed in a linear structure (in projection). The RGZ DR1 catalogue is a catalogue of radio source classifications that matches up all related radio components to the host galaxy observed in the $W1$ infrared image. In this study, we used the catalogue of cross-matched host galaxies as our primary input sample.

\subsection{Data formatting and cross-matching}
\label{sec:format}

Pre-processing was applied to both the MiraBest and RGZ~DR1 data-sets following the approach described in \citet{Aniyan2017} and \citet{Tang2019}. For the MiraBest data-set this has been described previously in \cite{Bowles2021} and \cite{Scaife2021}. In this work we create a machine learning data-set from the RGZ~DR1 catalogue with equivalent pre-processing. 

For each object in the RGZ catalogue with 15 < LAS < 270\,arcseconds, the catalogued RA and declination were used to query the FIRST survey using the Skyview Python API. From these search parameters the postage stamp server returns postage stamp images with $300\times 300$ pixels. The FIRST pixel size is 1.8\,arcsec. 

The following pre-processing steps are then performed:
\begin{enumerate}
\item Crop the central $150\times 150$ pixels.

\item All pixels outside a radial distance of $0.6\times$ the LAS in pixels were set to zero;

\item All NaN-valued pixels were set to zero;

\item We use the "outermost level" parameter from the RGZ~DR1 catalogue to implement an amplitude thresholding. This is equivalent to $3\times \sigma_{\rm rms, local}$ and we set all pixels with values below this level to zero.

\item Finally we normalise the image using,
\begin{equation}
    \text{Output} = 255\cdot\frac{\text{Input} - \text{min}(\text{Input})}{\text{max}(\text{Input})-\text{min}({\text{Input}})}.
\label{eqn:pre_processing_norm}
\end{equation}
where the minimum of the input is defined by the "outermost level" parameter. 

\end{enumerate}

The final images with $150\times150$ pixels have a size equivalent to 270\,arcseconds (4.5\,arcmin;  0.075\,degrees), which is the limiting LAS for selection from the RGZ catalogue.

RGZ DR1 entries were cross-matched with the catalogue of Miraghaei \& Best \citep[][ MiraBest]{Miraghaei2017} using the Vizier Python API. Based on the right ascension and declination  of a source in RGZ DR1, matches were requested within a radius of $0.5\times$ the LAS, i.\,e. if an entry from the MiraBest catalogue was found within the unmasked FOV then it was considered a match. The presence of a match (1) or no match (0) was recorded in the meta-data for each data sample in the RGZ DR1 machine learning data-set. Matched data samples were discarded from the dataset in order to ensure that there was no overlap between the labelled MiraBest subset, $\mX_{\rm l}$, and the unlabelled RGZ pool, $\mX_{\rm u}$.

\section{Choice of SSL Algorithm}
\label{sec:fixmatch}

\begin{figure*} 
    \centering
    \includegraphics[width=0.8\linewidth]{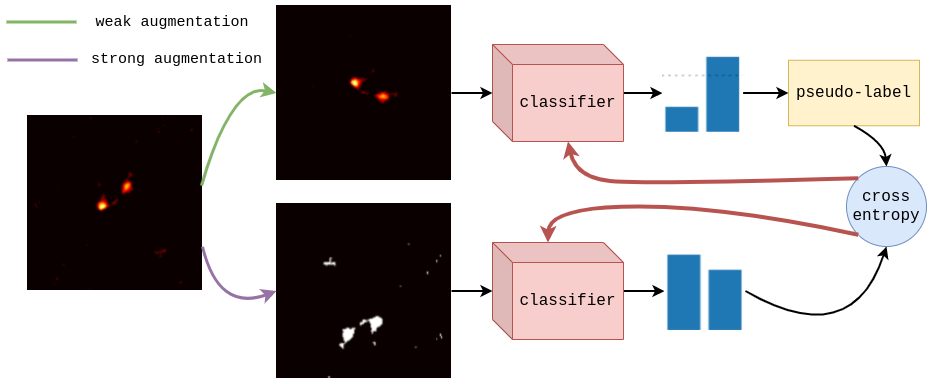}
    \caption{An unlabelled data point flowing through the FixMatch algorithm \citep{Sohn2020FixMatch:Confidence}. One strongly augmented and one weakly augmented image is generated. The classifier makes a prediction on each, if the softmax value of the weakly augmented prediction is above a threshold value $\tau$, consistentcy between the predictions on the two augmentations is enforced via the FixMatch loss.}
    \label{fig:fixmatch}
\end{figure*}

We selected the FixMatch algorithm \citep{Sohn2020FixMatch:Confidence} from the pool of SSL techniques as it has been shown to achieve state of the art performance on benchmarking data-sets, has relatively few hyperparameters, and is computationally cheap. FixMatch has already been successfully applied across a broad range of domains, such as action recognition in videos \citep[][]{Singh2021Semi-SupervisedLearning}, audio classification \citep[][]{Grollmisch2021ImprovingFixMatch} and image classification \citep{Sohn2020FixMatch:Confidence}.

FixMatch makes use of unlabelled data through consistency regularisation and pseudo-labelling by adding a loss term computed on two different augmentations of the same image. 
The "weak" augmentation, denoted $\alpha(\cdot)$, retains the semantic meaning of the image and simply uses the typical rotations and flipping augmentations applied as standard in most deep learning classification implementations. The "strong" augmentation, denoted by $\mathcal{A}( \cdot) $,  may significantly alter the image, by applying a sequence of augmentations that do not necessarily preserve the semantic meaning of the image. 

After augmentation, a \emph{pseudo-label} is generated by assigning a label if the classifier softmax output is greater than a threshold, $\tau$, for the weakly augmented image. This pseudo-label is then used as a target to compute the cross-entropy of the model's class prediction for the strongly augmented image and it is the corresponding "pseudo" cross-entropy that is minimised by the the optimiser to train the model. The full form of the pseudo-loss term is given by
\begin{equation}
\label{eq:fmuloss}
\mathcal{L}_{\rm u} = \sum_{u=0}^{\mu B}\underbrace{\mathds{1}(\textup{max}(p_{\rm m}(\alpha(\vx_{\rm u}))) \geq \tau)}_\text{threshold mask} \times \underbrace{H(\hat{q}_{\rm u}, p_{\rm m}(y | \mathcal{A}(\vx_{\rm u})))}_\text{pseudo-label cross-entropy},
\end{equation}
where $\mathds{1}$ is the indicator function and $\hat{q_{\rm u}}$ is the one-hot pseudo-label prediction on the weakly augmented image. $B$ is the labelled batch size, $\mu$ and $\tau$ are hyperparameters controlling the unlabelled batch size and confidence threshold, respectively; $p_{\rm m}$ is the softmax output of the classifier. Figure~\ref{fig:fixmatch} illustrates graphically how an unlabelled data point flows through the model to update the gradients of the model. 

The pseudo-loss term, $\mathcal{L}_{\rm u}$, is linearly combined with the supervised cross-entropy loss on the true labelled data using a tuning parameter, $\lambda_{\rm u}$, to give the complete FixMatch loss function:
\begin{equation}
    \mathcal{L} = \lambda_{\rm u} \mathcal{L}_{\rm u} + \sum_{l=0}^B \underbrace{H(y_{\rm l} , p_m(y | \alpha(\vx_{\rm l})))}_\text{supervised cross-entropy loss},
    \label{eq:fixmatchloss}
\end{equation}
where in this work we use the canonical value of $\lambda_{\rm u} = 1$.

\subsection{Model Architecture}
\label{sec:arch}

To provide a realistic baseline, we use the convolutional architecture from \citet{Tang2019}, which has high accuracy for radio galaxy classification and comparatively few ($\sim250$\,k) parameters which helps avoid overfitting at low data volumes. The network has 5 convolutional layers with ReLU activation and batch normalisation for each layer, with 3 max pooling layers in total to reduce the dimensionality. This is followed by 3 fully connected layers, each with ReLU activation and a dropout layer. A softmax layer at the end squeezes predictions between 0 and 1. We refer to this architecture as the ``Tang network'' (details can be found in \citet{Tang2019}).

\subsection{Data augmentations}
\label{subsec:dataprep}

\subsubsection{Weak augmentations}
\label{sec:weakaug}

We use the same weak augmentation for all algorithms. All pixel values in $\mX_{\rm l}$ and $\mX_{\rm u}$  are scaled to a mean of zero and a variance of one, calculated on $\mX_{\rm l} \cup \mX_{\rm u}$. Radio galaxy class is expected to be equivariant to orientation and chirality \citep["handedness"; see e.\,g.][]{ntwaetsile2021,Scaife2021} and we therefore rotate and flip each sample from $\mX_{\rm l}$ and $\mX_{\rm u}$ to a random orientation on ingest to the model.

\subsubsection{Strong augmentations}
\label{sec:strongaug}

We adjust the RandAugment algorithm \citep{Cubuk2019} to generate strong augmentations that only include transformations valid on black and white images. RandAugment applies $N$ sequential transformations with a given strength, $m = \textup{randint}(0, M)$, where $M \in [0, 10]$ and larger values of $m$ denote stronger distortions. Each sequential augmentation has a probability $p_{\rm aug}$ of occurring. We set $N=2$, $M=10$, $p_{\rm aug}=1$, consistent with \cite{Sohn2020FixMatch:Confidence}.

Table~\ref{tab:strongaug} shows a list of the strong augmentations used in this work, all of which can be found in the Python Imaging Library\footnote{\url{https://pillow.readthedocs.io} \label{fnote:pil}} or in our code. We note that although cropping is often used as a strong augmentation, we suggest that in this specific application cropping may be problematic as radio galaxy images are highly sparse (i.e. they contain a significant number of empty pixels), and that a randomly centred (upscaled) crop will mostly result in images of zeros. We found in our experiments that cropping significantly damages model performance and for this reason we omit it from the list of transformations passed to RandAugment. 

\begin{table*}
    \centering
    \caption{Strong augmentations used in this work. The table shows the name of each augmentation on the left, followed by a short description of the augmentation on the right. \label{tab:strongaug}}
    \begin{tabular}{r|l}
        \hline
        \textbf{Name} & \textbf{Description} \\\hline
        \textbf{AutoContrast} & Cuts off some of the darkest/brightest pixels in the image histogram and then remaps the brightest/darkest pixels to white/black. \\
        \textbf{Brightness} & Increases the pixel values of the image. \\
        \textbf{Contrast} & Increase/decrease image contrast depending on input parameter. \\
        \textbf{Identity} & No change to the image.\\
        \textbf{Sharpness} & Blur/sharpen the image depending on input parameter.\\
        \textbf{Equalize} & Forces grey scale pixels into a uniform distribution. \\
        \textbf{Posterize} & Reduces the number of possible magnitudes pixels can take by binning the pixel magnitudes. \\
        \textbf{ShearX} & Distort the image in the $x$ direction by shifting half of the image to the right and half to the left. \\
        \textbf{ShearY} & Distort the image in the $y$ direction by shifting half of the image upwards and half downwards.\\
        \textbf{Solarize} & Invert all pixels above a given brightness threshold.\\
        \textbf{SolarizeAdd} & Increase image brightness by a constant and then solarize. \\
        \textbf{TranslateX} & Translate the image in the $x$ direction.\\
        \textbf{TranslateY} & Translate the image in the $y$ direction.\\\hline
    \end{tabular}
\end{table*}

\subsection{Unresolved sources in the RGZ DR1 data-set}
\label{sec:cut}

\begin{figure}
\centering
\subfloat[A random sample of unresolved sources from the RGZ data. Images are cropped to 95x95.]{%
    \includegraphics[width=0.8\columnwidth]{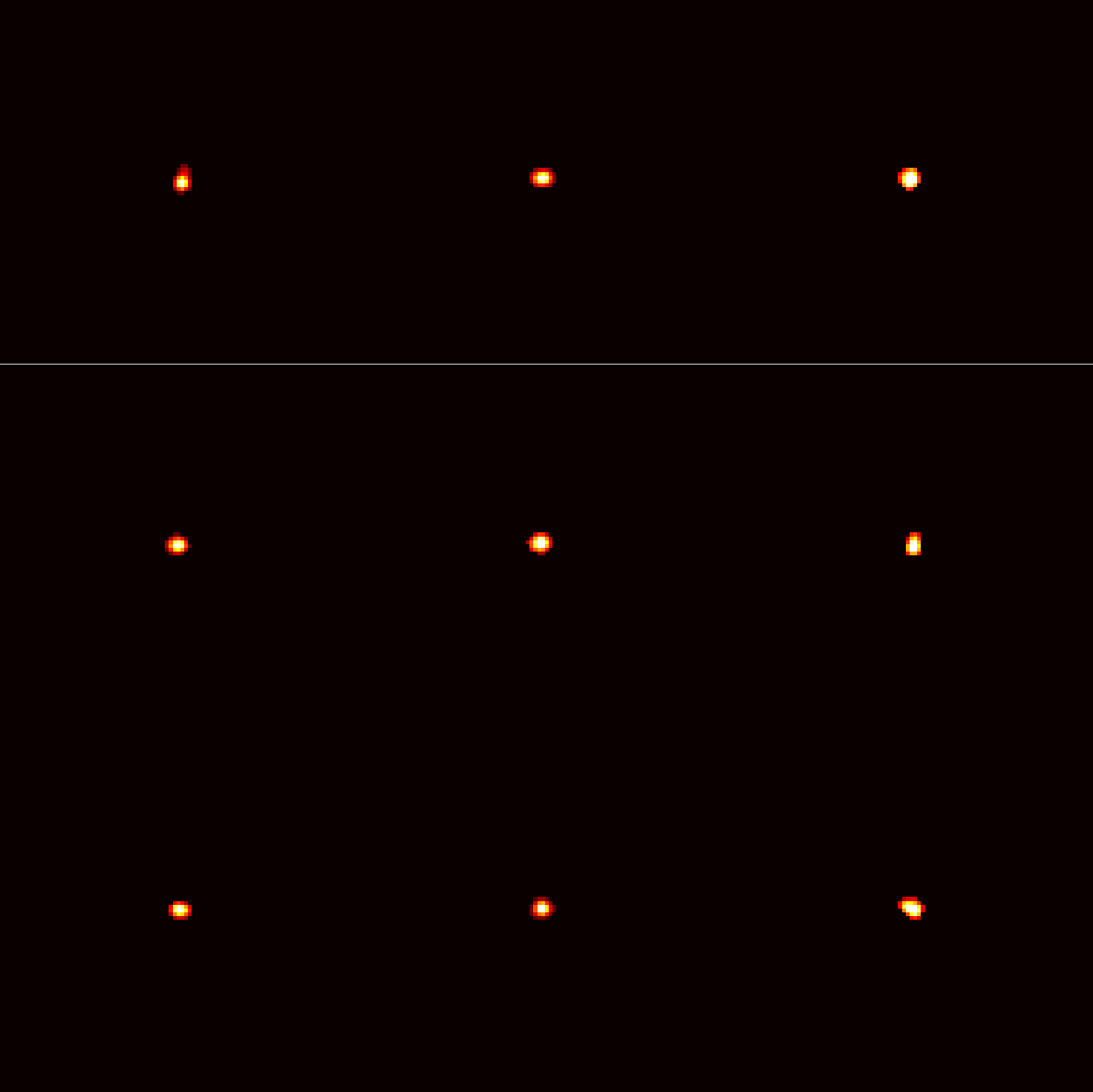}
    \label{fig:unresolved}
}

    \subfloat[A random sample of RGZ data with an angular size of 28 arsec and above. Images are cropped to 95x95.]{%
    \includegraphics[width=0.8\columnwidth]{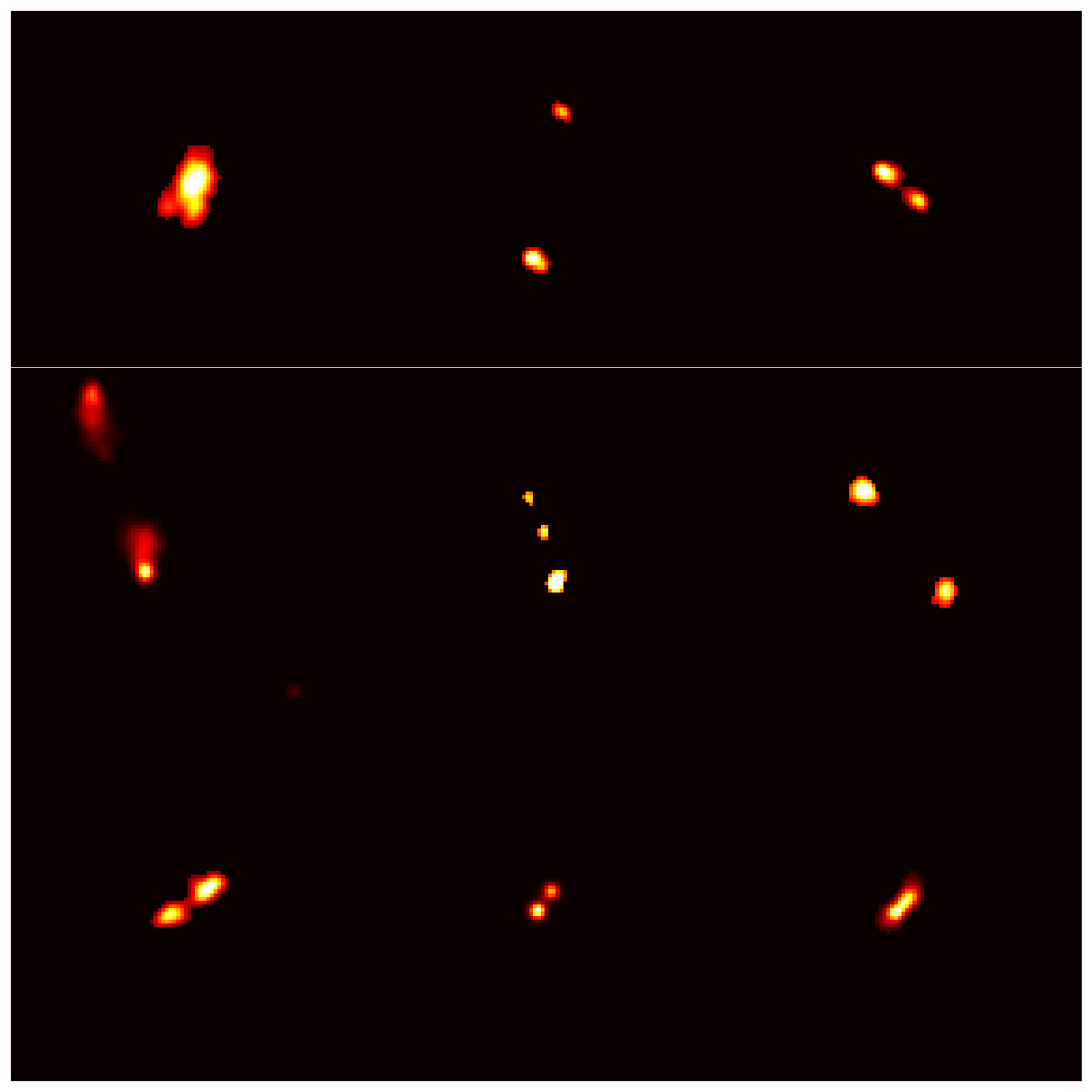}
    \label{fig:resolved}
    }
    \caption{Randomly selected samples from the RGZ DR1 data-set at different cut thresholds.}
\label{fig:cuts}
\end{figure}

\begin{figure*}
    \centering
    \includegraphics[width=0.8\linewidth]{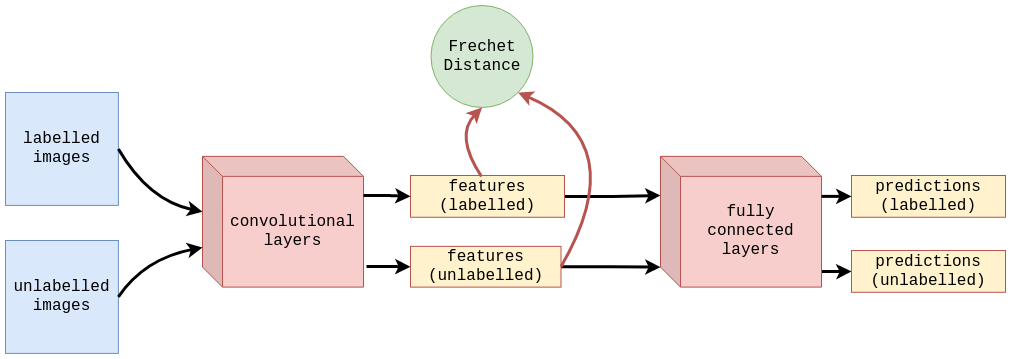}
    \caption{Schematic showing how to extract feature representations for Frechet Distance calculation. A supervised classifier is trained on all the labels. The (un)labelled data-set is passed through the classifier and the output of the convolutional layers is eased as a representation for each data-set. The similarity of the representations is computed using the Frechet distance.}
    \label{fig:fdschematic}
\end{figure*}

\begin{figure}
    \centering
    \includegraphics[width=0.4\textwidth]{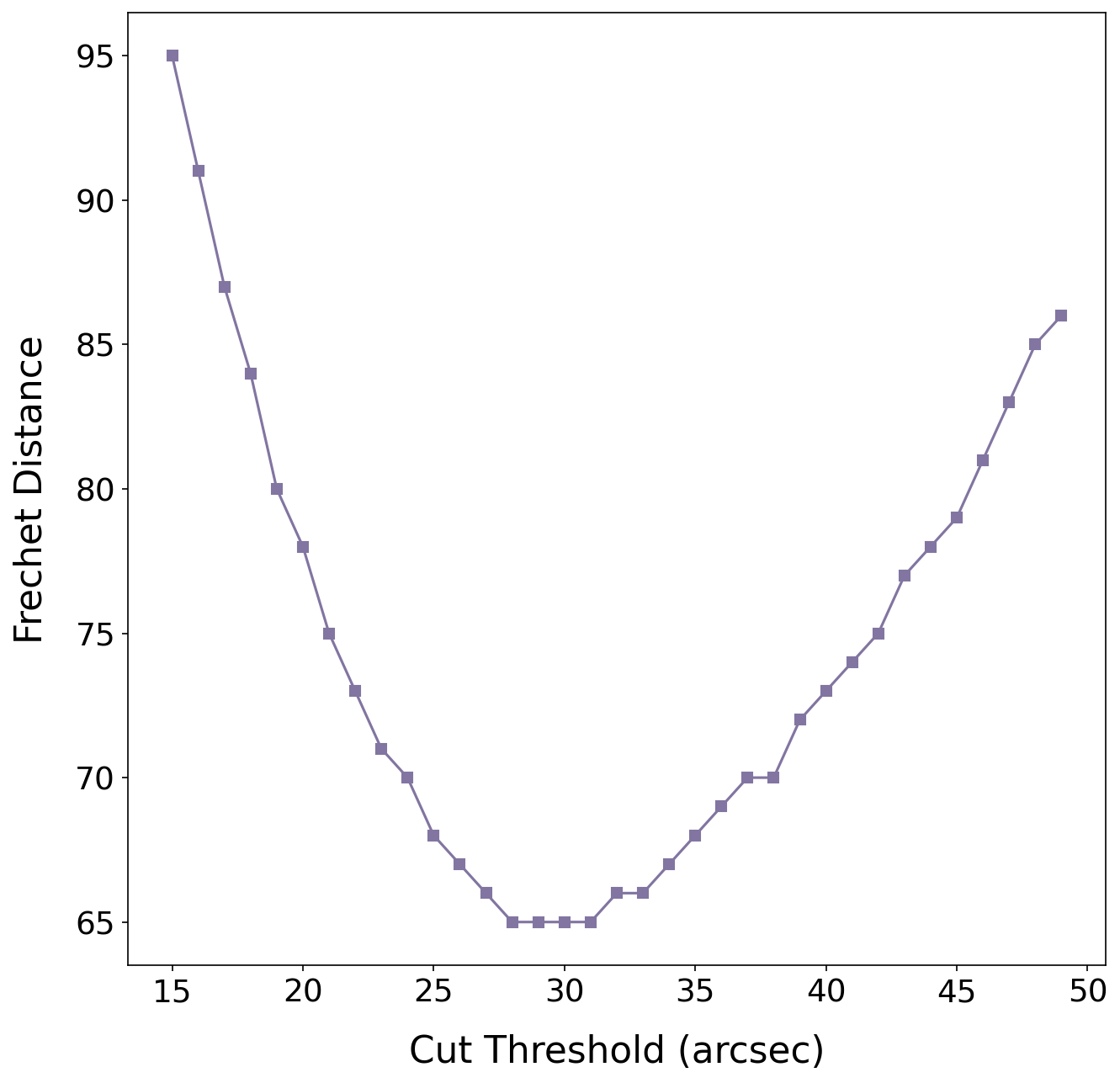}
    \caption{Frechet distance between the MiraBest and RGZ DR1 data-sets as a function of different angular size cut thresholds, see Section~\ref{sec:cut} for details.}
    \label{fig:dshiftFID}
\end{figure}



While the RGZ~DR1 data originate from the same radio survey and are pre-processed in the same manner as MiraBest, the choice of filters for selecting data-points in the RGZ~DR1 case is significantly wider ($\sim 10^5$ datapoints) and its true class balance is unknown. Furthermore, we find that the RGZ DR1 data-set contains many unresolved sources, see Figure~\ref{fig:unresolved}, which can overwhelm the learning signal. We remove these unresolved sources by enforcing a hard lower threshold on source extension. To calculate the exact value of this threshold, we estimate the similarity of the RGZ DR1 and MiraBest data-sets over a range of threshold values and select the cut which gives the most similar data-sets. 

As a proxy for similarity we use a modification of the Frechet Inception Distance \citep[FID;][]{Heusel2017} between MiraBest and RGZ DR1. The FID is a metric commonly used to evaluate the similarity of synthetic data-sets to the training set for generative algorithms, such as when training GANs \citep{Goodfellow2014}, and has previously been adapted for use in radio astronomy when evaluating the synthetic radio galaxy populations generated from the latent space of a variational autoencoder \citep{Bastien2021}. To calculate this quantity, we first train the Tang network on the MiraBest data-set using all available labels. We then separately forward pass both the RGZ DR1, denoted $\mX_{\rm RGZ}$, and MiraBest data-set, denoted $\mX_{\rm MB}$, through the trained classifier, ignoring the classification outputs and instead taking the feature representations for each data-set, denoted $\mF_{\rm RGZ}$ and $\mF_{\rm MB}$, respectively. The feature representations are simply the flattened output of the final convolutional layer of the Tang network (after activation, batchnorm and pooling). Figure~\ref{fig:fdschematic} illustrates this process. These feature representations can be interpreted as a low dimensional embedding of each dataset. We assume the features to follow a multidimensional Gaussian probability distribution for each dataset, allowing us to calculate the mean, $\vmu$, and covariance matrix, $\mC$, for both data-sets. The Frechet Distance (FD) between the two data-sets (in feature space) is then given by
\begin{equation}    
\begin{aligned}
    \label{eq:fid}
    {\rm FD}(\mX_{\rm RGZ}, \mX_{\rm MB}) = & ||\vmu_{\rm RGZ} - \vmu_{\rm MB}||^2_2 \: \: + \\
    &  \Tr \left( \mC_{\rm MB} + \mC_{\rm RGZ} - 2(\mC_{\rm RGZ} \mC_{\rm MB})^{\tfrac{1}{2}} \right),
\end{aligned}
\end{equation}
which gives us a similarity metric between the RGZ DR1 and MiraBest data-sets, with lower FD values indicating a higher level of similarity.

In Figure~\ref{fig:dshiftFID} it can be seen that there is an obvious minimum in the FD where the two data-sets are most similar. As there is a small plateau in FD at the minimum, we use the lowest cut threshold that gives the minimum FD, equal to $\theta_{\rm cut} = 28\,$arcsec, in order to remove the smallest number of samples. Examples of data samples from the MiraBest dataset with angular sizes above this threshold are shown in Figure~\ref{fig:resolved}.

\subsection{Model training}
\label{sec:training}

We split the available data into validation and training sets by choosing randomly stratified  splits (i.\,e. class balance is preserved). The training data is used to optimise the model weights and the validation set is used to optimise the hyperparameters of the model and to choose the best performing weights through a method known as early stopping.

We keep the validation set size realistic by scaling it to $20 \%$ of $\mX_l$ with a hard lower limit of 53 data points (5\% of the total MiraBest training data) to produce meaningful results when $\mX_l$ is small, as an extremely small validation set causes the validation loss/accuracy to be too noisy, making it difficult to select a good model. We set the learning rate to 0.005 but scale the batch size, $B$, with $\mX_l$. Astronomical data-sets vary in size and content: large validation sets may not be available. Therefore, our models need to perform well across a range of scenarios without brittleness to small cahnges in hyperparameters. For these reasons, we keep hyperparameter tuning to a minimum, opting instead to choose reasonable values close to the optimal values in the computer vision literature \citep{Sohn2020FixMatch:Confidence}. We choose $\mu = 7$, which allows the model to use a large proportion of the unlabelled data in a single epoch, and $\tau = 0.95$ which allows the model's confidence on unlabelled data to pass the threshold before it begins to overfit. We allow a greater number of maximum training epochs when training using FixMatch as the model is less prone to over-fitting and  requires a greater number of epochs to result in softmax predictions above the threshold, $\tau$, for the unlabelled data. Early stopping is used to choose the weights with the highest validation accuracy, which we empirically find to give better performance than using the validation loss. 

Fair evaluation is a problem in the SSL literature \citep{Oliver2018RealisticAlgorithms}, yet is crucial if we wish to apply SSL methods to real data. To ensure reproducibility and to evaluate the variance in our results, each experiment is averaged over 10 runs initialised using the random seeds 0-9. This ensures consistent training and validation data splits during each experiment. A single test set is used to estimate the model's generalisability to unseen data after choosing the model weights with the best validation set accuracy.

Experiments were performed on a single Nvidia A100 GPU with a total of 28 days of runtime which includes all realisations of random seedings required for statistically significant results in some of the experiments. We used Weights \& Biases \citep{wandb} to track experiment results.

\section{Model Performance}
\label{sec:results}
\subsection{Baseline results}
\label{sec:baseline}

\begin{table*}
    \centering
    \caption{Baseline FRI/II classification results on the MiraBest dataset with different numbers of labels. Values given are means of 10 aggregated runs with uncertainties given by the standard error.}
    \label{table:baseline}
    \begin{tabular}{llllllllll}
        \hline
        \# of labels & Accuracy (\%) & Label & Precision & Recall & F1 & \\
        
        \midrule
        \multirow{2}{*}{10}   &  \multirow{2}{*}{$60.9 \pm 2.7$}  & FRI  &  $0.588 \pm 0.027$ &  $0.647 \pm 0.057 $ & $0.608 \pm 0.035 $ & \\ 
        {} & {} &  FRII &   $ 0.759 \pm 0.018 $ &  $ 0.731 \pm 0.026$  &  $0.741 \pm 0.014$ & \\ 
        \vspace{1mm} \\ 

        \multirow{2}{*}{50}   &  \multirow{2}{*}{$76.2 \pm 1.7$}  & FRI  &  $ 0.774 \pm 0.025 $ &  $ 0.728 \pm 0.027 $ & $ 0.747 \pm 0.020 $ & \\ 
        {} & {} &  FRII &   $ 0.760 \pm 0.016$ &  $0.794  \pm 0.028 $  &  $ 0.774 \pm 0.017$ & \\  
        \vspace{1sp} \\ 
        
        \multirow{2}{*}{101}   &  \multirow{2}{*}{$ 81.6 \pm 1.2$ }  & FRI  &  $0.812 \pm 0.020$ &  $0.818 \pm 0.023$ & $0.811 \pm 0.012$ &  \\ 
        {} & {} &  FRII &   $0.832  \pm 0.016 $ &  $ 0.815 \pm 0.027 $  &  $0.819 \pm 0.014$ & \\  
        \vspace{1sp} \\ 
        
        \multirow{2}{*}{203}   &  \multirow{2}{*}{$84.1 \pm 0.7$}  & FRI  &  $0.865 \pm 0.018 $ &  $0.804 \pm 0.021$ & $0.830 \pm 0.008$ &\\ 
        {} & {} &  FRII &   $0.830 \pm 0.014$ &  $0.876  \pm 0.020 $  &  $0.850 \pm 0.008$ & \\  
        \vspace{1sp} \\ 
        
        \multirow{2}{*}{301}   &  \multirow{2}{*}{$84.4 \pm 0.6$}  & FRI  &  $0.847 \pm 0.001 $ &  $0.830 \pm 0.008$ & $0.838 \pm 0.006$ & \\ 
        {} & {} &  FRII &   $0.844 \pm 0.006$ &  $0.858  \pm 0.010 $  &  $0.851 \pm 0.005$ & \\  
        \vspace{1sp} \\ 

        \multirow{2}{*}{393}   &  \multirow{2}{*}{$86.4 \pm 0.6$}  & FRI  &  $0.872 \pm 0.007 $ &  $0.843 \pm 0.013 $ & $0.857 \pm 0.007$ & \\ 
        {} & {} &  FRII &   $0.859 \pm 0.010$ &  $0.884  \pm 0.007 $  &  $0.871 \pm 0.006$ & \\  
        \vspace{1sp} \\ 
        
        \multirow{2}{*}{855}   &  \multirow{2}{*}{$86.9 \pm 0.5$}  & FRI  &  $0.891 \pm 0.004 $ &  $0.831 \pm 0.010$ & $0.860 \pm 0.007$ & \\ 
        {} & {} &  FRII &   $0.852 \pm 0.008$ &  $0.905  \pm 0.004 $  &  $0.877 \pm 0.005$ & \\  

        \bottomrule
    \end{tabular}
    
    \end{table*}

To inform our decision making about analysis pipelines for future surveys, it is important to give a fair comparison with the current techniques used. This will help decision making when trading off increased complexity and computational overhead with performance for future pipeline design, as we expect our results on archival data to hold given new data-sets in the same domain.

We establish a baseline model performance using the Tang network, see Section~\ref{sec:arch}, by training in a fully supervised fashion with the data-set weakly augmented, see Section~\ref{sec:weakaug}, and the results of this baseline are shown in Table~\ref{table:baseline}. As expected, it can be seen that baseline classification performance increases with the number of labelled samples. We see a steep drop off at low label volumes and a gentler increase at higher label volumes as the model approaches the accuracy with all labels.

Our baseline is in line with state of the art performance in the literature for the Mirabest data-set (including \emph{Uncertain} samples), such as \citet{Scaife2021} who achieve $85.30 \% \pm 1.35 \%$ accuracy by using an equivariant CNN, and \citet{Bowles2021} who achieve $84 \% $ accuracy with an attention gated CNN. We also note that the baseline achieves comparable accuracy to the fully labelled case with only 393 labels, indicating that the data-set has significant redundancy in information.


\subsection{Discarding labels to create an artificial SSL scenario (Case A)}
\label{subsec:fmmb}

\begin{table*}
    \centering
    \caption{FixMatch FRI/II (case A) classification results with different numbers of labels. Values given are means of 10 aggregated runs with uncertainties given by the standard error.}
    \label{table:fixmatch}
    \begin{tabular}{lllllllll}
        \hline
        \# of labels & Accuracy (\%) & Label & Precision & Recall & F1  & \\
        
        \midrule
        \multirow{2}{*}{10}   &  \multirow{2}{*}{$60.2 \pm 2.7$}  & FRI  &  $0.584 \pm 0.029$ &  $0.703 \pm 0.032 $ & $0.631 \pm 0.021 $ &\\ 
        {} & {} &  FRII &   $ 0.638 \pm 0.024 $ &  $ 0.5078 \pm 0.060$  &  $ 0.551 \pm 0.047$ & \\ 
        \vspace{1mm} \\ 

        \multirow{2}{*}{50}   &  \multirow{2}{*}{$80.8 \pm 1.6$}  & FRI  &  $ 0.803 \pm 0.022 $ &  $ 0.805 \pm 0.022 $ & $ 0.802 \pm 0.017 $ &\\ 
        {} & {} &  FRII &   $ 0.819 \pm 0.017$ &  $0.810  \pm 0.024 $  &  $ 0.812 \pm 0.016$ & \\  
        \vspace{1sp} \\ 
        
        \multirow{2}{*}{101}   &  \multirow{2}{*}{$ 83.3 \pm 1.3$ }  & FRI  &  $0.824 \pm 0.009$ &  $0.835 \pm 0.015$ & $0.828 \pm 0.006$ & \\ 
        {} & {} &  FRII &   $0.846  \pm 0.011 $ &  $ 0.830 \pm 0.013 $  &  $0.837\pm 0.005$ & \\  
        \vspace{1sp} \\ 
        
        \multirow{2}{*}{203}   &  \multirow{2}{*}{$85.1 \pm 1.1$}  & FRI  &  $0.869 \pm 0.015 $ &  $0.820 \pm 0.028$ & $0.840 \pm 0.014$ & \\ 
        {} & {} &  FRII &   $0.845 \pm 0.019$ &  $0.880  \pm 0.018 $  &  $0.859 \pm 0.010$ & \\  
        \vspace{1sp} \\ 
        
        \multirow{2}{*}{301}   &  \multirow{2}{*}{$84.9 \pm 0.8$}  & FRI  &  $0.853 \pm 0.008 $ &  $0.832 \pm 0.016$ & $0.842 \pm 0.009$ & \\ 
        {} & {} &  FRII &   $0.848 \pm 0.012$ &  $0.865  \pm 0.009 $  &  $0.856 \pm 0.007$ & \\  
        \vspace{1sp} \\ 

        \multirow{2}{*}{393}   &  \multirow{2}{*}{$84.9 \pm 0.5$}  & FRI  &  $0.869 \pm 0.010 $ &  $0.8122 \pm 0.017 $ & $0.838 \pm 0.007$ &  \\ 
        {} & {} &  FRII &   $0.836 \pm 0.011$ &  $0.884  \pm 0.011 $  &  $0.858 \pm 0.004$ & \\  
        \vspace{1sp} \\ 
        
        \bottomrule
    \end{tabular}
    
    \label{table:fmmb}
    \end{table*}

In the SSL literature, algorithms are tested by throwing away labels from a labelled data-set to create ``unlabelled'' data. Although this approach is not realistic for a true use case in radio astronomy, it is important that we first isolate the effect of using an astronomical data-set rather than a computer vision data-set before testing on real unlabelled data. This gives us the most similar scenario to those explored in the SSL literature. To do this, we recreate the same artificial scenario by throwing away labels from the MiraBest dataset. This allows us to set a best case upper bound on possible performance when using unlabelled data as we might naively expect the algorithm to perform worse or at best equivalently well to this scenario when using truly unlabelled data.

\begin{figure}
\centering
\subfloat[Accuracy on the test set.]{%
  \includegraphics[clip,width=0.8\columnwidth]{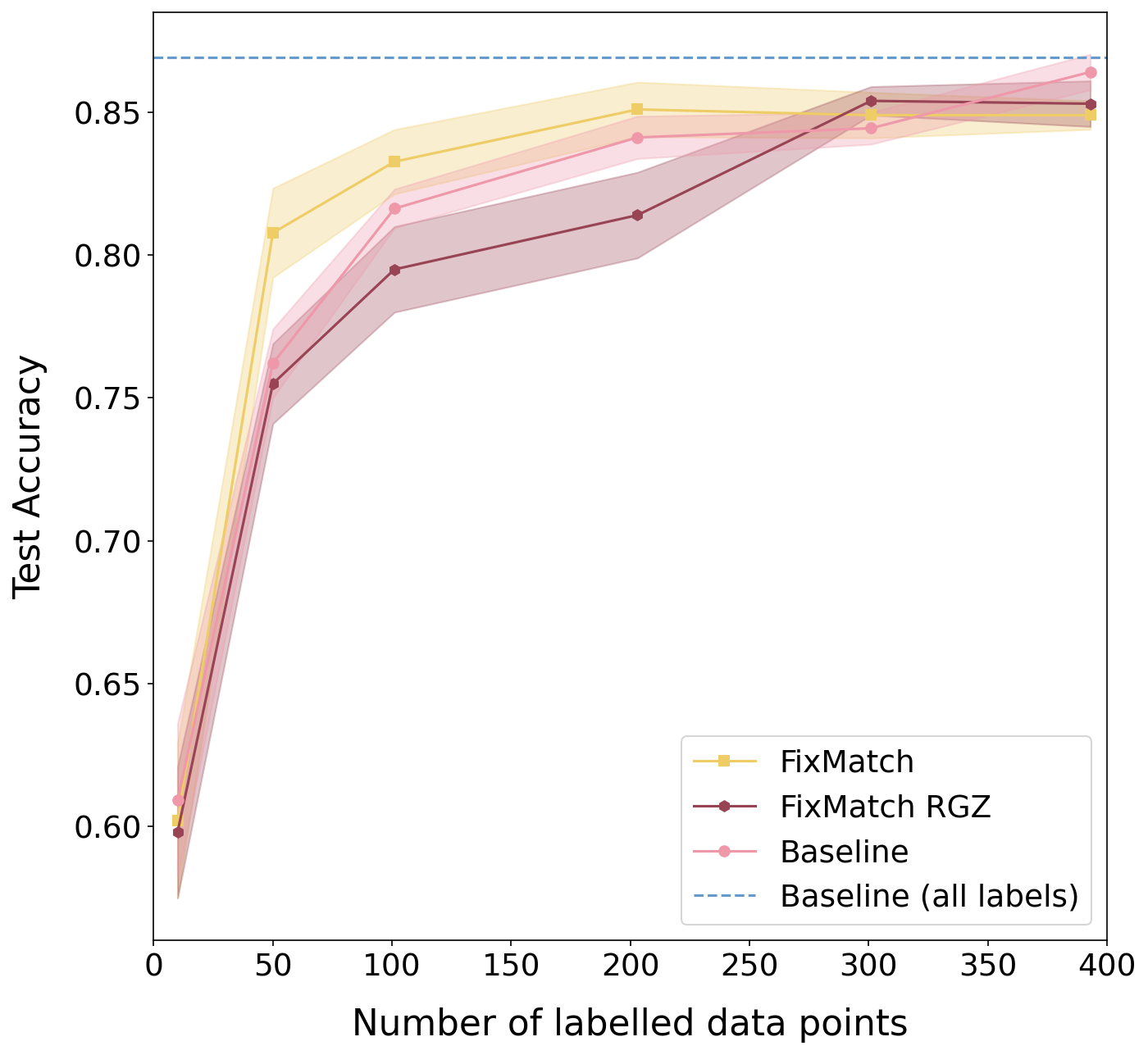}%
  \label{fig:datafracacc}
}

\subfloat[Loss on the test set.]{%
  \includegraphics[clip,width=0.8\columnwidth]{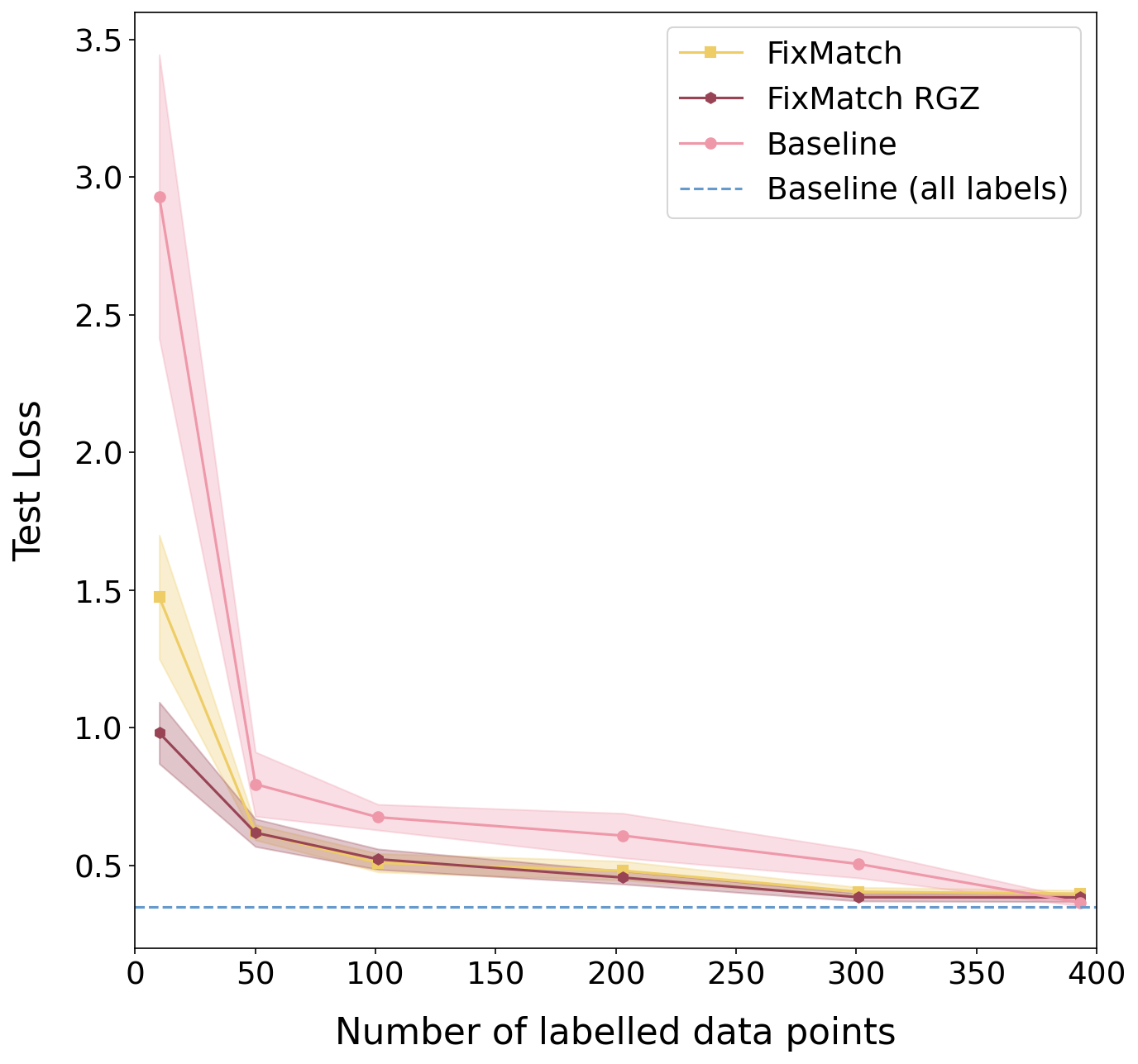}%
  \label{fig:datafracloss}
}
\caption{Model performance as a function of labelled dataset size for Case~A, see Section~\ref{subsec:fmmb} for details. Top: test accuracy; Bottom: test loss. $R$ is the ratio of unlabelled to labelled data, and the performance of a fully supervised model is shown in each case as a dashed line.}
\label{fig:datafrac}
\end{figure}

We use a small, randomly sampled subset of the MiraBest data-set \citep{Miraghaei2017} as $\mX_{\rm l}$ and the remainder as $\mX_{\rm u}$, equivalent to the case shown in Figure~\ref{subfig:sslset}. In all cases we keep the data stratified. Table~\ref{table:fmmb} gives a comprehensive overview of our results. In Figure~\ref{fig:datafrac} we plot the test set accuracy and loss of Case~A against the baseline model for different label volumes. It can be seen that the use of FixMatch in Case~A achieves a consistently lower loss on the test data and outperforms the baseline, see Section~\ref{sec:baseline}, in test set accuracy when there are 203 labels or fewer. 

We are able to recover comparable accuracy ($85.10 \pm 1.13\%$) to the supervised baseline with all labels ($86.93 \pm 0.54 \% $) using just $20\%$ (203) of the labels. However, FixMatch's performance degrades quickly with fewer than 50 labels, which is in stark contrast to similar experiments performed by  \cite{Sohn2020FixMatch:Confidence} on more standard benchmark datasets, where good performance was achieved even with only one sample per class. However, we note that this "sweet spot" where FixMatch has a significant advantage over the baseline model does not cover the full range of $R$ (ratio of unlabelled to labelled data) and is quite narrow.

\begin{figure}
\centering
\subfloat[Validation loss with 50 labelled samples.]{%
  \includegraphics[clip,width=0.9\columnwidth]{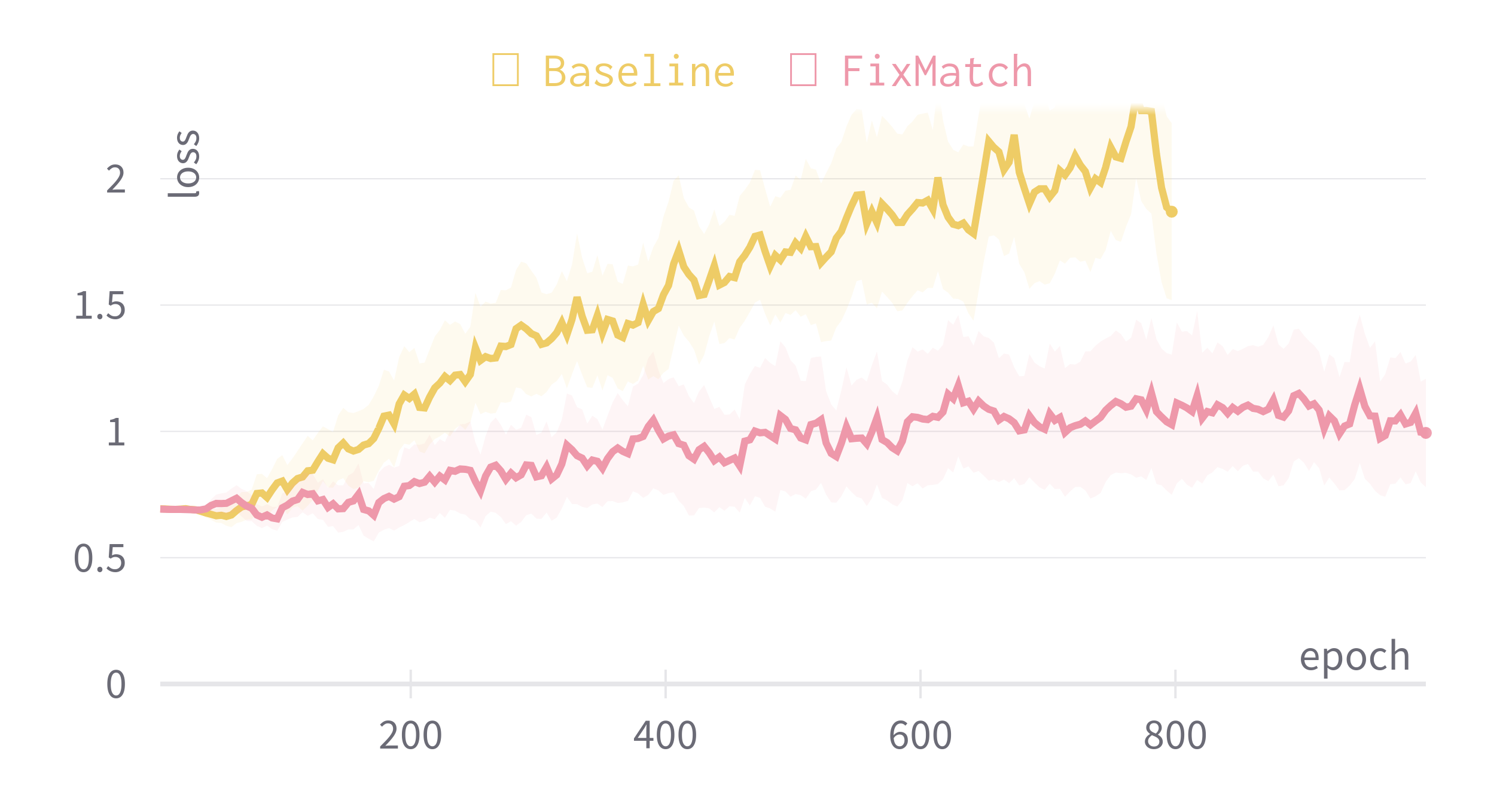}%
  \label{fig:val_loss50}
}

\subfloat[Validation loss with 393 labelled samples.]{%
  \includegraphics[clip,width=0.9\columnwidth]{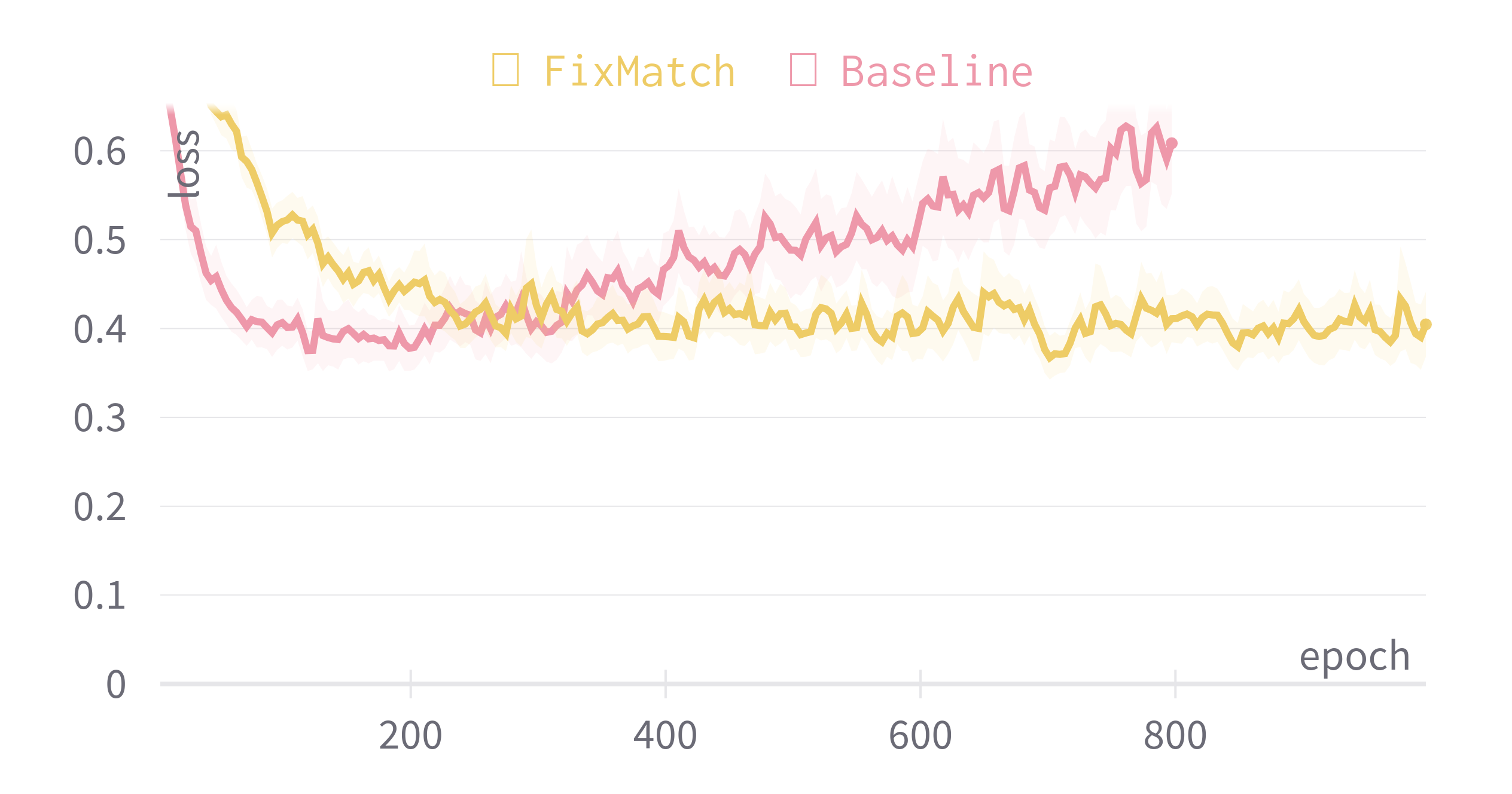}%
   \label{fig:val_loss393}
}
\caption{FixMatch regularisation effect on validation loss. The shaded area shows the standard error over 10 runs. We use exponential moving average smoothing with a weight of 0.3.}
\label{fig:val_loss}
\end{figure}


\label{sec:datatest}

\subsection{Testing FixMatch on real unlabelled data (Case B)}
\label{subsec:fmrgz}

\begin{table*}
    \centering
    \caption{FixMatch FRI/II (case B) classification results with different numbers of labels. Values given are means of 10 aggregated runs with uncertainties given by the standard error.}
    \label{table:fixmatchrgz}
    \begin{tabular}{lllllllll}
        \hline
        \# of labels & Accuracy (\%) & Label & Precision & Recall & F1  & \\
        
        \midrule
        \multirow{2}{*}{10}   &  \multirow{2}{*}{$59.8 \pm 2.3$}  & FRI  &  $0.580 \pm 0.024$ &  $0.697 \pm 0.048 $ & $0.622 \pm 0.022 $ &\\ 
        {} & {} &  FRII &   $ 0.644 \pm 0.027 $ &  $ 0.505 \pm 0.061$  &  $ 0.545 \pm 0.005$ & \\ 
        \vspace{1mm} \\ 

        \multirow{2}{*}{50}   &  \multirow{2}{*}{$ 75.5 \pm 1.4$}  & FRI  &  $ 0.760 \pm 0.019 $ & $0.734 \pm 0.035 $ & $ 0.740 \pm 0.020 $ &\\ 
        {} & {} &  FRII &   $ 0.764 \pm 0.018$ &  $0.775  \pm 0.030 $  &  $ 0.764 \pm 0.015$ & \\  
        \vspace{1sp} \\ 
        
        \multirow{2}{*}{101}   &  \multirow{2}{*}{$79.5 \pm 1.5$ }  & FRI  &  $0.792 \pm 0.025$ &  $0.799 \pm 0.019$ & $0.791 \pm 0.013$ & \\ 
        {} & {} &  FRII &   $0.810  \pm 0.013 $ &  $ 0.792 \pm 0.031 $  &  $0.798 \pm 0.017$ & \\  
        \vspace{1sp} \\ 
        
        \multirow{2}{*}{203}   &  \multirow{2}{*}{$81.4 \pm 1.5$}  & FRI  &  $0.825 \pm 0.027 $ &  $0.803 \pm 0.026$ & $0.807 \pm 0.013$ & \\ 
        {} & {} &  FRII &   $0.825 \pm 0.018$ &  $0.824  \pm 0.041 $  &  $0.816 \pm 0.02$ & \\  
        \vspace{1sp} \\ 
        
        \multirow{2}{*}{301}   &  \multirow{2}{*}{$85.4 \pm 0.5$}  & FRI  &  $0.876 \pm 0.016 $ &  $0.818 \pm 0.015$ & $0.843 \pm 0.006$ & \\ 
        {} & {} &  FRII &   $0.841 \pm 0.010$ &  $0.887  \pm 0.018 $  &  $0.862 \pm 0.006$ & \\  
        \vspace{1sp} \\ 

        \multirow{2}{*}{393}   &  \multirow{2}{*}{$85.3 \pm 0.8$}  & FRI  &  $0.865 \pm 0.011 $ &  $0.827 \pm 0.018 $ & $0.844 \pm 0.010$ &  \\ 
        {} & {} &  FRII &  $0.846 \pm 0.013$ &  $0.877  \pm 0.011 $  &  $0.861 \pm 0.008$ & \\  
        \vspace{1sp} \\ 
        
        \bottomrule
    \end{tabular}
    
    \label{table:fmrgz}
    \end{table*}

When training our model in a real scenario, we do not know what our unlabelled data contains. The model is unlikely to perform as well as in the ``ideal SSL'' Case A scenario, so we test how close we can get when using real unlabelled data. This will give a better indication of whether the algorithm is robust enough for general use in radio astronomy, and if not, how much domain specific development is required. This will help decision making as to whether FixMatch is a useful stream of research for future pipeline development.

We train our model using FixMatch with a large pool of 20\,000 unlabelled data samples from RGZ~DR1, before applying a cut threshold as discussed in Section~\ref{sec:cut}, with labelled samples from MiraBest. This is equivalent to the case shown in Figure~\ref{subfig:domainexpansion}, where the unlabelled samples have some covariate shift from the labelled and test samples as a result of the selection biases inherent in the creation of the MiraBest catalogue.

In Figure~\ref{fig:datafrac} we see an almost identical improvement (decrease) in test set loss as in Case A (see Table~\ref{table:fmrgz} for all results). However, this does not translate to an improvement in accuracy - the model actually performs consistently \textit{worse} than the baseline on our test set.


\subsection{Model Calibration}
\label{sec:calibration}

\begin{figure}
    \centering
    \includegraphics[width=0.4\textwidth]{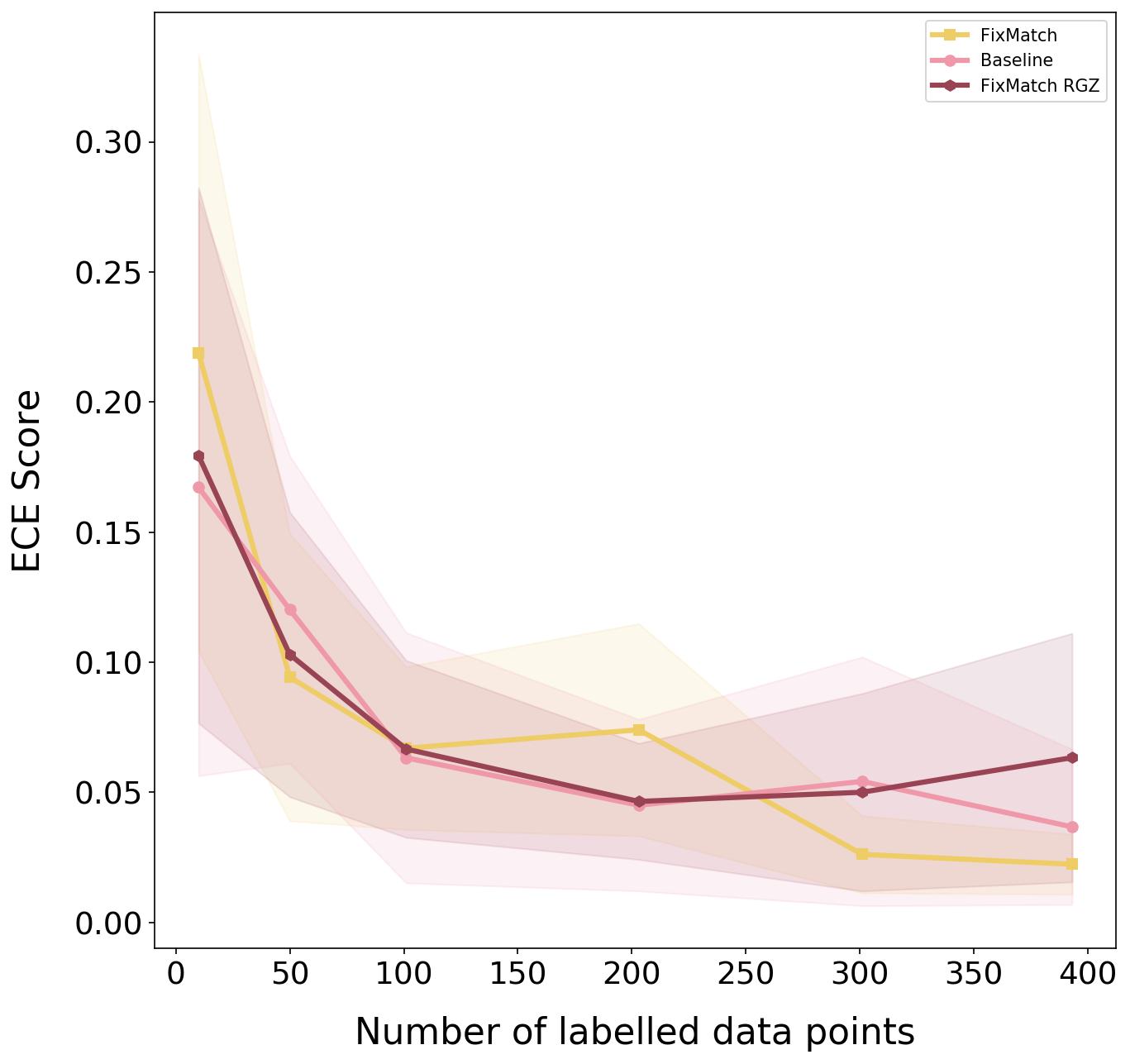}
    \caption{ECE scores with different numbers of labels. Values given are means of 10 aggregated runs with uncertainties given by the standard error.}
    \label{fig:ece}
\end{figure}

In order to perform science with our models, it is important that we can interpret our model output in a meaningful way. It is tempting to view softmax outputs as true probabilities, allowing us to find anomalous or out of distribution sources by simply looking at data with a low softmax output. However, the only constraint placed on the softmax outputs is that they must sum to 1 - there is no pressure from the loss function to calibrate softmax values to a probabilistic confidence. We can quantify how well calibrated the model is by using the Expected Calibration Error \citep[ECE;][]{Guo2017OnNetworks}, which we slightly modify. 

We bin the test set softmax outputs of the model into 10 bins, using a non-constant bin size such that we have the same number of data points in each bin. We calculate the error rate, $\kappa_b$, of the predictions (i.\,e. an estimate for the true probabilistic accuracy of the model) and the average of the softmax probabilities, $\mu_b$, for each bin, $b \in B$. The ECE is then given by the average of these differences, with a lower ECE indicating better calibration:
\begin{equation}
\label{eq:ece}
    {\rm ECE} = \sum_{b \in B} \frac{| \kappa_b - \mu_b |}{|B|}.
\end{equation}

We calculate the ECE at different data volumes for the baseline, Case~A and Case~B. Figure~\ref{fig:ece} shows our results, where we see a steep improvement in calibration as the number of samples initially increases for all algorithms. However, above $\sim$100 labels, the benefit of extra labels on calibration becomes much smaller. Furthermore, we do not see any benefit from using FixMatch on model calibration.

\section{Discussion}
\label{sec:disc}

\subsection{Does FixMatch outperform the baseline?}
\label{sec:compbase}

For our model to be useful, it needs to significantly outperform the baseline to justify the extra hyperparameters, complexity and computational overhead. In Section~\ref{subsec:fmmb} we see that FixMatch consistently outperforms the baseline when training with 50-200 labels, achieving significantly higher test set accuracy and lower loss. We hypothesise that this is both due to a regularisation effect from the strongly augmented samples, as well as the model learning new information from the unlabelled data. This is illustrated in Figure~\ref{fig:val_loss50}, where the well-behaved validation loss during training with FixMatch demonstrates its robustness to overfitting. We also observe that this effect is not as pronounced with 393 labelled samples, which is reflected in the equal minimum test loss in Figure~\ref{fig:val_loss393}, implying that in the high data (label) limit, there is less benefit to using FixMatch.

In Section~\ref{subsec:fmrgz}, we still see the same regularisation effect as in Case~A, as shown by the almost identical loss curves in Figure~\ref{fig:datafracloss} for both Case~A and B. This suggests that the regularisation effect of FixMatch is decoupled from the algorithm's ability to learn from the unlabelled data. However, Case~B is unable to achieve better-than-baseline classification accuracy, regardless of the number of labels. We believe this is due to the covariate shift between the MiraBest data-set, which comprises the labelled data and test set, and the unlabelled RGZ DR1 data-set. The model may be learning to make better predictions on data from the RGZ DR1 distribution, whereas we are testing the model's performance on a test set drawn from the MiraBest data-set, therefore our evaluation is skewed towards models that perform well on $\mX_{\rm test}$ rather than on truly unseen data. This means that although our model does not perform as well on our test set, it is unclear whether the opposite would be true if we our test set was drawn from the RGZ DR1 catalogue.

Since the two data-sets are preprocessed identically and drawn from the same survey, the only other sources of covariate shift are the different filters and cuts used for each data-set, see Section~\ref{sec:data} for details. This can manifest itself in a number of ways, such as unseen sub-populations or class imbalance in the unlabelled data. 

\subsection{Effect of class imbalance in the unlabelled data-set on model performance.}
\label{subsec:classimb}

\begin{figure}
    \centering
    \includegraphics[width=0.4\textwidth]{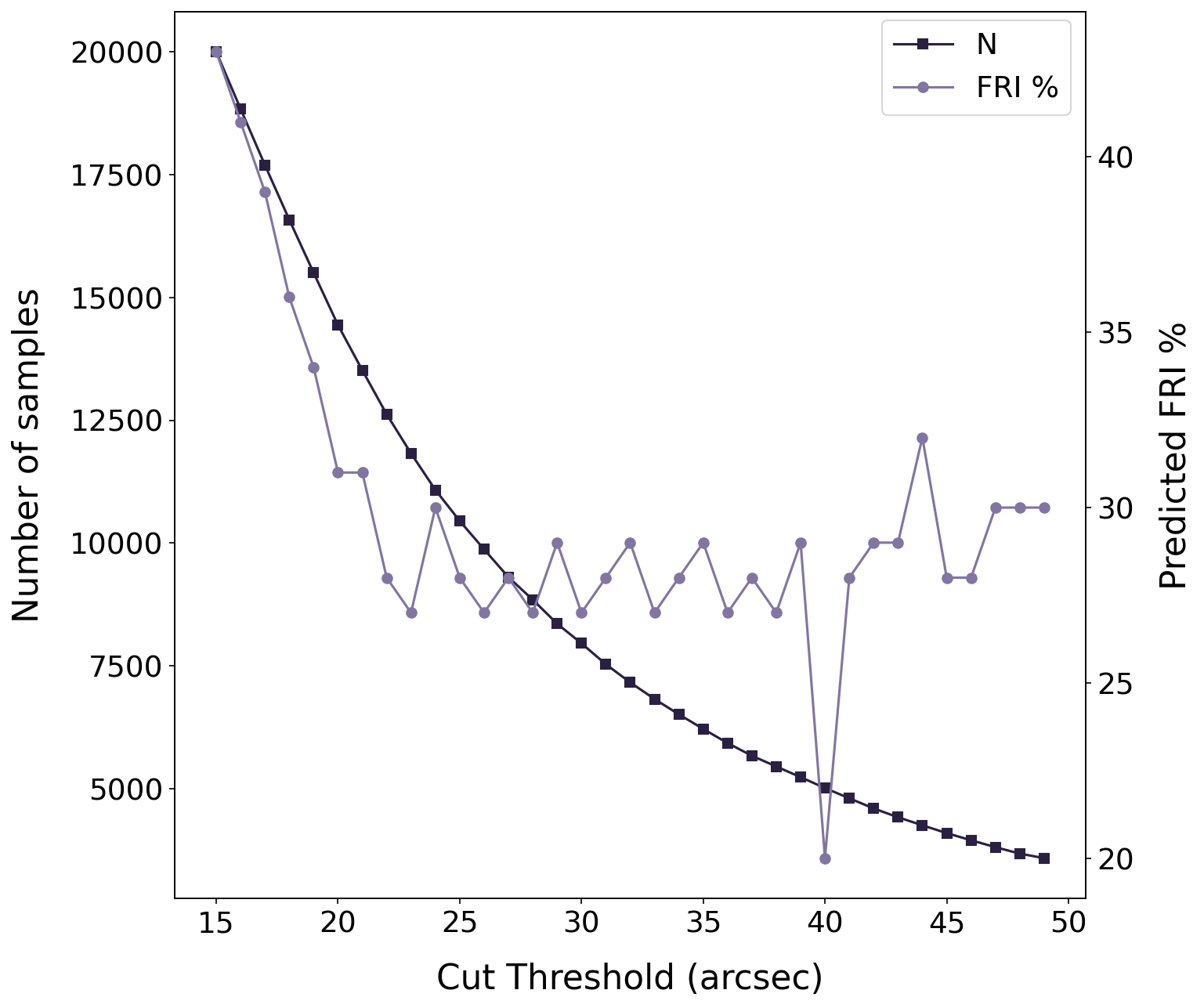}
    \caption{Number of samples remaining in the RGZ DR1 dataset and predicted proportion of FRI type objects as a function of angular size cut threshold.}
    \label{fig:dshiftNFR}
\end{figure}

\begin{figure}
\centering
\subfloat[Accuracy on the test set after removing FRI samples from the unlabelled data-set. Pearson R coefficient: 0.914.]{%
  \includegraphics[clip,width=0.8\columnwidth]{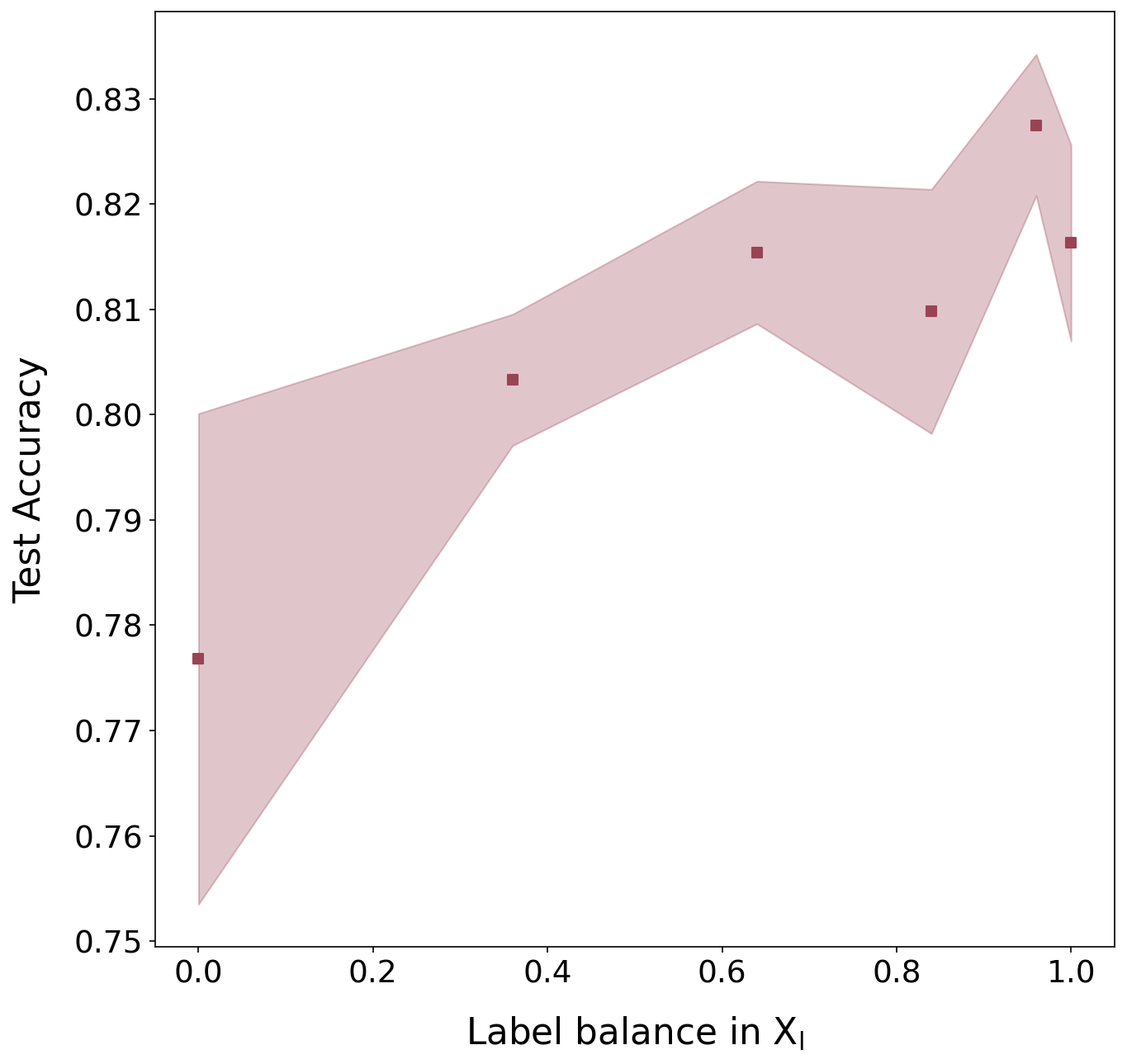}%
  \label{fig:balaccfri}
 }

 \subfloat[Accuracy on the test set after removing FRII samples from the unlabelled data-set. Pearson R coefficient: 0.852.]{%
   \includegraphics[clip,width=0.8\columnwidth]{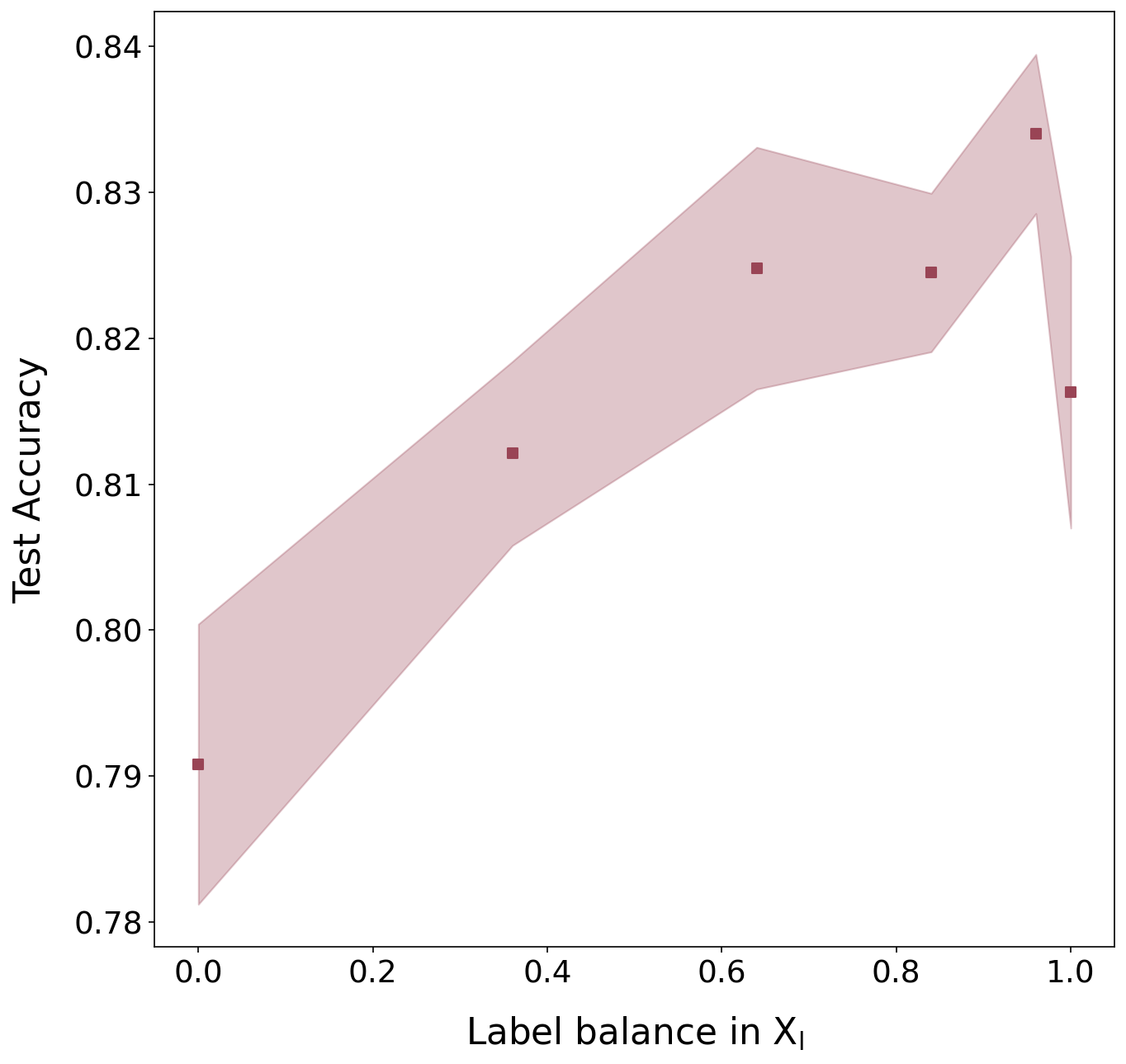}%
   \label{fig:balaccfrii}
 }
\caption{Accuracy as a function of label balance in the unlabelled (MiraBest) data-set. We only use Confident data in the labelled subset (69 labels) to reduce the noise in our results. Error bars show the standard error after aggregating 20 runs (seeded 0-19).}
\label{fig:balacc}
\end{figure}

One possible cause of dataset shift is label imbalance, more properly referred to as prior probability shift \citep{datasetshiftbook}. This is a commonly occurring problem in classification of astronomical data-sets, \citep[e.\,g. photometric classification;][]{Boone2019Avocado:Augmentation}. Since we cannot know the class balance of unlabelled astronomical data, it is important that model performance is not brittle to skewed unlabelled data class balance. Furthermore, this type of failure may be difficult to diagnose once the model has been deployed, as we cannot check the unlabelled data for imbalance. However, while we cannot know the exact class balance of $\mX_{\rm u}$, we can make an estimate by using an accurate classifier and looking at its predictions when tested on those unlabelled data. Figure~\ref{fig:dshiftNFR} shows the proportion of FRIs predicted by the baseline classifier trained using all labelled data and tested on the RGZ DR1 data-set. The figure indicates these data likely have a high class imbalance regardless of cut threshold choice. 

We can test whether the class imbalance in the RGZ~DR1 dataset is causing a difference in model performance due to prior probability shift by repeating Case A, but introducing an artificial imbalance into the unlabelled data pool and measuring the effect of this imbalance on test set accuracy.

To examine whether this can fully explain the performance drop described in Section~\ref{subsec:fmrgz} we train a model on the MiraBest data-set (Case A) with 69 labelled data samples and artificially imbalance the remaining unlabelled MiraBest samples by removing FRI/II samples until the unlabelled data-set contains $\beta$ FRI samples as a proportion of the total unlabelled samples. We then quantify label balance by computing $4(1-\beta) \beta $, which normalises the value between 0 and 1 and makes it symmetric with respect to FRI or FRII imbalance. We perform this test using only the data qualified as \emph{Confident} in the MiraBest dataset in order to reduce noise in the results introduced by the higher predictive uncertainty associated with the \emph{Uncertain} samples.

Figure~\ref{fig:balacc} shows the test set accuracy as a function unlabelled class balance when removing either FRI or FRII samples from $\mX_{\rm u}$. It can be seen that there is a clear trend in classifier test accuracy, which improves for more balanced $\mX_{ \rm u}$, indicating that class imbalance in $\mX_{\rm u}$ is an important factor in the performance of the model when tested on $\mX_{\rm test}$. Pearson R values of 0.914 and 0.852 show that the correlation between label balance and test set accuracy is significant. These results suggest that class imbalance in the RGZ data-set may be the cause of the poorer accuracy for Case~B reported in Section~\ref{subsec:fmrgz}.


\subsection{Effect of labelled subset choice on model performance}
\label{sec:subsets}

\begin{table}
    \centering
    \caption{Test set accuracy using 50 labelled samples from different MiraBest categories (Case~A, see Section~\ref{subsec:fmmb} for details). The uncertainties given are the standard error calculated over 10 runs.}
    \label{tab:categories}
    \begin{tabular}{llllll}
        Algorithm & All (\%) & Confident (\%) & Uncertain (\%)\\
        \midrule
        Baseline &  $76.21 \pm 1.67$  &  $81.90 \pm 1.26$ & $68.5 \pm 2.13$\\  
        FixMatch & $80.78 \pm  1.56$ & $\bm{82.29} \pm 0.85$ &  $ 68.82 \pm 2.56$\\ 
        \vspace{0.5mm}
    \end{tabular}
\end{table}

During the experiments in Section~\ref{sec:datatest}, different data splits, produced using from different random seeds, resulted in large discrepancies in test loss accuracy. For example, with 50 labelled samples there is a difference of $\sim16\%$ between the best and worst runs, implying that some data splits within the MiraBest dataset are significantly more informative to the model than others. This may be due to the structure of the MiraBest data-set, which is split into \emph{Confident} and \emph{Uncertain} categories, which indicate the human labeller's confidence in the label, see Section~\ref{sec:mirabest}.

Understanding which labels are most informative to the model is an important goal, as it can inform which data samples to label to optimise the effectiveness of future models with the lowest labelling cost. To test this, we isolate training datasets from the \emph{Confident} and \emph{Uncertain} sub-populations to test which one of these is more informative to our model, or whether it is beneficial to use both. 

We run experiments by training on 50 labelled samples from either the \emph{Confident} subset, \emph{Uncertain} subset, or a random sample of both (equivalent to Case~A, see Section~\ref{subsec:fmmb}). The validation set is sampled from the same subset as the labelled data. In all cases, unlabelled data (if used) are randomly sampled from both \emph{Confident} and \emph{Uncertain} subsets. The results are shown in Table~\ref{tab:categories} where it can be seen that using only \emph{Uncertain} labelled data greatly degrades model performance, decreasing accuracy by over $23 \%$. Using \emph{Confident} samples alone gives the best performance, which is perhaps surprising given that the test set contains both \emph{Confident} and \emph{Uncertain} data samples,  
and may suggest that those samples with more ambiguous labelling, class overlap or incorrect labels cause the classifier to learn a more disturbed decision function. This effect might potentially dominate any gains from the training samples spanning more of the data manifold, reducing accuracy of the final classifier. Given this result, we believe that the variance in performance seen when using both \emph{Confident} and \emph{Uncertain} is predominantly due to the variance in the number of \emph{Confident} samples chosen for the labelled dataset.



\subsection{Can we use Frechet Distance as a proxy for model performance?}





\begin{figure}
    \centering
    \includegraphics[width=0.4\textwidth]{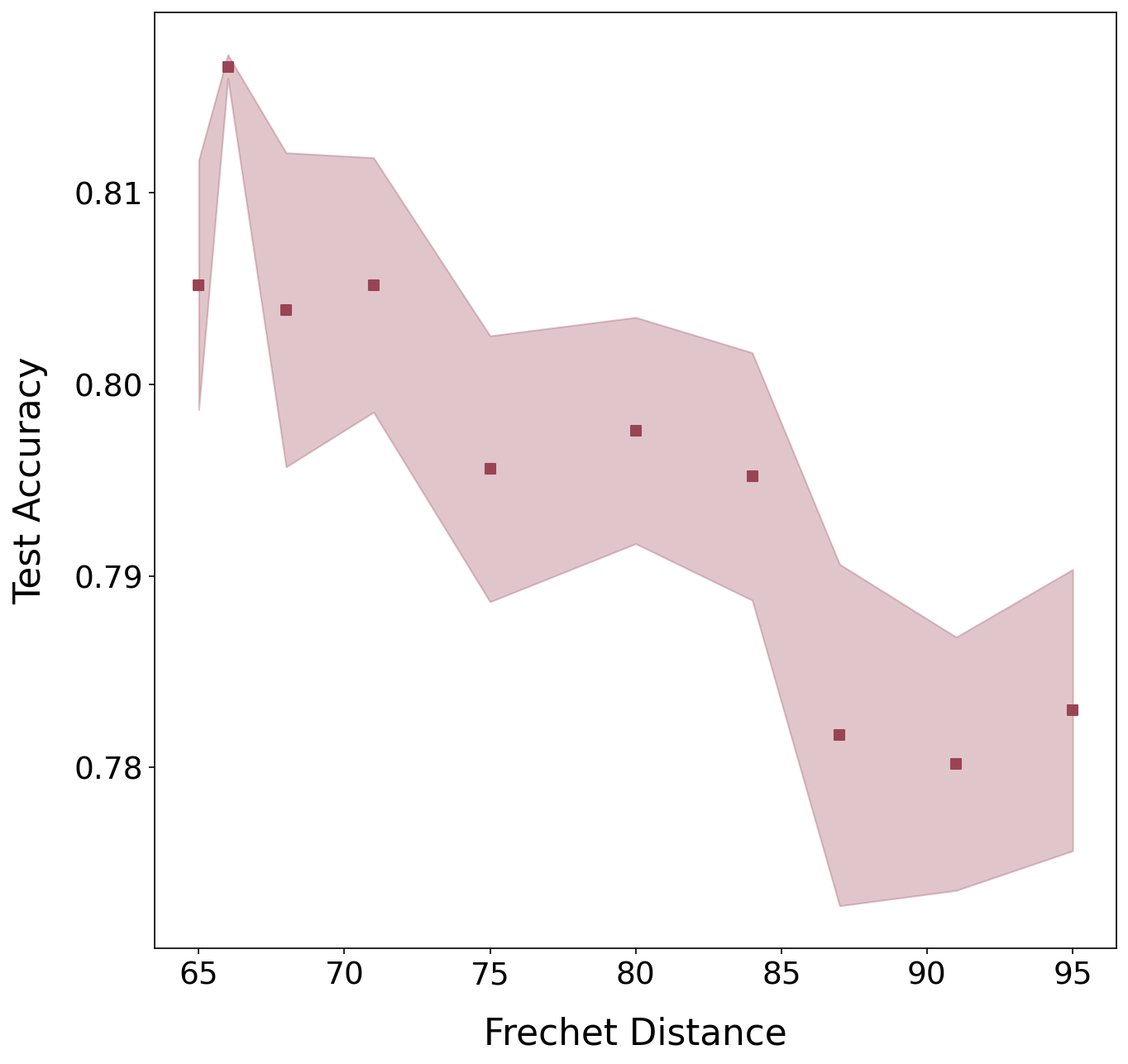}
    \caption{Test set accuracy of FixMatch (case B) as a function of Frechet Distance between RGZ and MiraBest data-sets. The shaded area shows the standard error calculated over 30 runs. Pearson R coefficient: 0.920}
    \label{fig:fidacc}
\end{figure}

In general, testing models through repeated training and examination of test set performance is computationally expensive. Furthermore, FixMatch has a significant computational overhead compared with the baseline, as many more data-points need to be passed both forwards and backwards through the model. It would therefore be useful to have a predictor of the performance for the SSL model (FixMatch) that is fast to compute. This would allow us to efficiently test the viability of different unlabelled data-sets. 

One reason for poor SSL performance that is particularly relevant for our use case is covariate shift between the labelled and unlabelled data. Therefore, if we can measure the covariate shift in a computationally cheap manner, we may be able to predict whether an SSL technique is suitable for the problem. Here we do this using the Frechet Distance between $\mX_{\rm u}$ and $\mX_{\rm l}$, see Section~\ref{sec:cut}. 
To test the effect of this covariate shift this we run Case~B experiments and recover model performance as a function of cut threshold, which by proxy changes the Frechet Distance, see Figure~\ref{fig:dshiftFID}, introducing a variable degree of covariate shift. For these tests we train on 65 data samples qualified as \emph{Confident}, use the RGZ DR1 data-set as the unlabelled data pool, and measure accuracy on the reserved test set. We only include data-points to the left of the minimum (inclusive) in Figure~\ref{fig:dshiftFID} as there is no motivation to remove more data than necessary from the unlabelled data. 

The results of these tests shown in Figure~\ref{fig:fidacc} show a negative correlation between the FD and the test accuracy (Pearson R 0.920), indicating that we can use the FD as a predictor for model performance. However, we note that the results are quite noisy and in this case 30 runs were required to reduce the uncertainty bounds sufficiently enough for a significant result. Whilst the nature of this metric may seem a good proxy for covariate shift, the empirical nature of its calculation may cause it to be affected by other forms of sample-dependent dataset shift, including prior probability shift, which may partly cause the noisy results. We therefore suggest that future applications of this metric would be strengthened by combining with a separate analysis to quantify the label imbalance in the unlabelled data that could give a decoupled measure of the prior probability shift.
    
\section{Conclusion and Future Work}
\label{sec:conclusion}

In this work we have demonstrated that the use of semi-supervised learning, implemented using the FixMatch algorithm, provides some regularisation benefits to radio galaxy classification when learning with few labelled samples, mitigating the effect of overfitting. Furthermore, we show that the model is able to learn from unlabelled data and that we achieve better accuracy on the test set using the SSL approach than the baseline with fewer labels in Case~A. While this is encouraging, the improvement in accuracy is much smaller than for the standard benchmark computer vision data-sets on which the algorithm was initially tested. We also find in Section~\ref{sec:calibration} that using an SSL algorithm does not improve model calibration.

Poor results using the truly unlabelled RGZ~DR1 data in Case~B highlight an important obstacle to applying SSL ``in the wild'' on scientific observational data where $\mX_{\rm l}$ and $\mX_{\rm u}$ are unlikely to be drawn from identical distributions. We find in Section~\ref{subsec:classimb} that class imbalance in the unlabelled data-set is detrimental to classifier performance and this may account for the effect of prior probability shift between the unlabelled and labelled data on test-set accuracy. However, we also note that we are measuring performance on a test set which is distributed identically to the labelled data-set, whereas real data may be distributed more similarly to the unlabelled data. Future work will look at estimating accuracy on unlabelled data, which would give a better estimate of model performance on real data. Furthermore, the effect of mismatched brightness distributions between the labelled and unlabelled data might be investigated. Since this semi-supervised scheme has has not been designed for covariate shifted unlabelled data, we would expect that future improvements can push model performance towards the upper bound set by Case A (see Section~\ref{subsec:fmmb} for details).

In Section~\ref{sec:subsets} we see that the effect of a poorly chosen labelled data-set on model performance is significantly greater than any improvement from the SSL algorithm. This highlights the importance of choosing to label data points which will be useful for our models. This result indicates that we need to be careful when choosing which data to label, as a poor choice can result in sub-optimal performance. Active learning may be a suitable approach to begin solving this problem. It may also be useful for identifying samples with a high level of uncertainty during training, and mitigating the effect of propagating incorrect labels in these cases by correcting or removing these images.

In order to test our choice of unlabelled data-set in a computationally efficient way, we use the Frechet Distance to measure covariate shift between the unlabelled and labelled data. We find that this metric can predict test set accuracy, and that it might be used as an initial guide when comparing different unlabelled data-sets to narrow down a search. However, our results are noisy due to high sample variance and the metric does not directly measure label imbalance. We believe it would be best used as part of a fuller analysis which also takes into account prior probability shift (label imbalance).

We believe that a naive application of SSL, although it may outperform the baseline in some cases and provide useful regularisation, requires further domain specific development to provide a worthwhile advantage in the case of radio galaxy morphology classification. Although the FixMatch algorithm has minimal overhead when passing through labelled data, there is a significant computational cost in passing through the unlabelled data. Furthermore, future work will need to address the problem of biased sampling in our labelled data-sets and the covariate shift between the unlabelled and labelled data-set before the extra cost can be justified. There is also a limit on how much unlabelled data can be used in a semi-supervised way as we cannot keep increasing the $\mu$ parameter indefinitely. Unsupervised pretraining is one possible solution to these problems \citep[][]{Hendrycks2019UsingUncertainty}, although this will require significantly more compute and larger models. We note that our work runs parallel to pretraining and that FixMatch can be used as a final fine-tuning layer, as has been done previously for computer vision applications \citep{Cai2021, Kim2021}. Furthermore, pre-training will allow larger volumes of unlabelled data to be used than is possible with FixMatch. StyleMatch \cite{Zhou2021Semi-SupervisedStyleMatch} extends FixMatch to help with domain generalisation, which may also allow data from different surveys to be used and could also be a useful area of research. 

We suggest that radio astronomy specific development will likely be needed to achieve better results with both semi-supervised learning and (contrastive) pretraining, as has been successfully done for gravitational lens identification \citep{Stein2021}. Specifically, a custom suite of augmentations for radio galaxy images should be developed, guided by the computer science literature \citep{Tian2020WhatLearning, Cubuk2019Autoaugment:Data}. While there have been attempts to automate this process \citep[e.g.][]{Tamkin2020}, they are somewhat limited in the search space of possible augmentations, and do not include domain specific knowledge.  We would therefore propose that augmentations are designed ''by hand`` and tested for performance.

\section*{Acknowledgements}
We thank the anonymous reviewer whose comments improved this work.

IVS, AMS, MB \& MW gratefully acknowledge support from the UK Alan Turing Institute under grant reference EP/V030302/1. IVS gratefully acknowledges support from the Frankopan Foundation. HT gratefully acknowledges the support from the Shuimu Tsinghua Scholar Program of Tsinghua University. 

This work has been made possible by the participation of more than 12,000 volunteers in the Radio Galaxy Zoo Project. The data in this paper are the result of the efforts of the Radio Galaxy Zoo volunteers, without whom none of this work would be possible. Their efforts are individually acknowledged at \url{ http://rgzauthors.galaxyzoo.org}.

IVS acknowledges the usefulness of \url{https://github.com/kekmodel/FixMatch-pytorch} for implementations.

\section*{Data Availability}
Code for this paper can be found at \url{https://github.com/inigoval/fixmatch}.

The RGZ DR1 catalogue will be made publicly-available through Wong et al (2022; in preparation). This work makes use of the MiraBest machine learning dataset, which is publically available under a Creative Commons 4.0 license at \url{https://doi.org/10.5281/zenodo.4288837}.

\bibliography{references, references2}
\bibliographystyle{mnras}

\clearpage

\end{document}